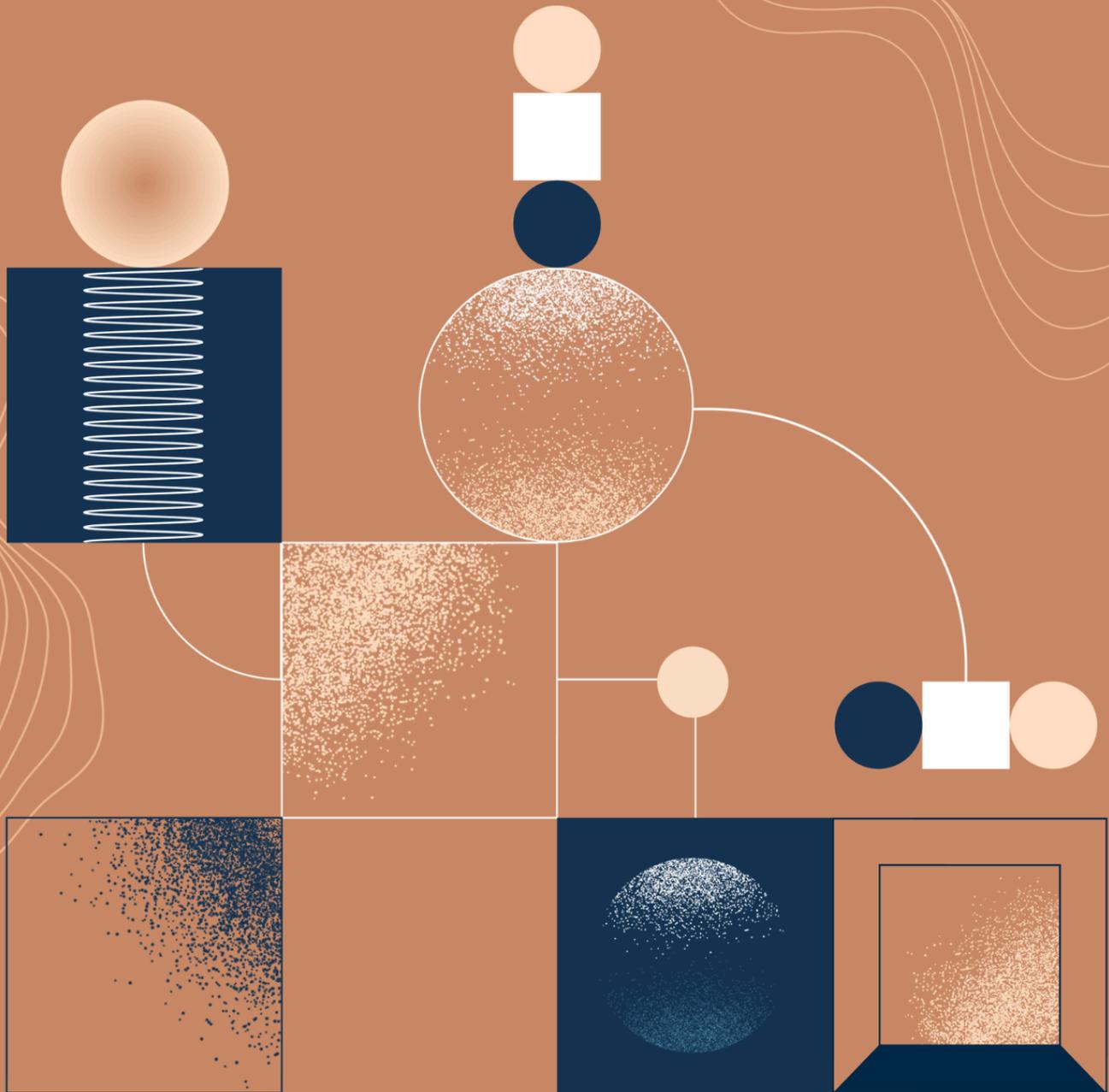

IAPS | Institute for AI Policy and Strategy

April 2025

# AI Agent Governance: A Field Guide

by: Jam Kraprayoon, Zoe Williams, and Rida Fayyaz

# Executive Summary

Agents—**AI systems that can autonomously achieve goals in the world, with little to no explicit human instruction about how to do so**—are a major focus of leading tech companies, AI start-ups, and investors. If these development efforts are successful, some industry leaders claim **we could soon see a world where millions or billions of agents are autonomously performing complex tasks across society. Society is largely unprepared for this development**.[1]

Today, the leading approach for developing agents leverages recent advances in foundation models like ChatGPT and Claude. Scaffolding software is built around these models which allow them to interact with various tools and services—enabling them to have long-term memory, plan and interact with other agents, and take actions in the world.

**While today's agents can do a variety of things**—from identifying critical vulnerabilities in software to ordering books on Amazon—**they still face serious limitations in completing more complex, open-ended, longer time-horizon tasks**.[2] Agents have major issues with reliability, as well as limitations in reasoning and digital tool use. There are also potential barriers to adoption if the processing power needed to run agents is cost-prohibitive. For example, newer AI systems that can 'think through' complex problems step-by-step (like the recently developed 'reasoning models') require much more processing power when answering questions or performing tasks, which can substantially drive up the energy and server costs needed to operate these systems.

Benchmarks designed to evaluate the performance of agents on real-world tasks consistently find that while current agents perform comparably to humans on shorter tasks, they tend to perform considerably worse than humans on tasks that would take an equivalent human one or more hours to complete.

---

[1] Meta's CEO, Mark Zuckerberg, told investors he wants to "introduce AI agents to billions of people" (Heath 2023) and Salesforce CEO Marc Benioff predicted there would be one billion AI agents in service by the end of FY2026 (Sozzi 2024).

[2] For an example of an agent identifying critical vulnerabilities in real-world code, see Google's Project Zero blog on Big Sleep (Project Zero 2024). For a visual demo of a browser agent being used to make an online shopping purchase, see this demo (AI Digest 2024).

Table 1: Agent performance on various benchmarks representing real-world tasks (as of December 2024)[3]

| Agent benchmark | Performance |
|---|---|
| General AI Assistants (GAIA) | Tests real-world assistant capabilities across personal tasks, science, and general knowledge. **Human accuracy (92%) far exceeds best agent performance (15%)**, with agents completely failing on complex multi-step tasks. |
| METR Autonomy Capability Evals | Evaluates skills in cybersecurity, software engineering, and machine learning. **Agents perform comparably to humans on tasks taking ~30 minutes, but complete less than 20% of tasks requiring 1+ hours of human time.** |
| RE-Bench | A benchmark for evaluating the AI agents' ability to automate the work of experienced AI R&D researchers. **Agents outperform humans in 2-hour tasks (4× better scores), but humans excel with longer timeframes—slightly better at 8 hours and doubling agent performance when given 32 hours.** |
| CyBench | Assesses cybersecurity capabilities through professional-level Capture the Flag challenges. **Agents struggled with tasks that take human teams more than 11 minutes to complete.** |
| SWE-bench Verified | Features real-world software engineering problems from GitHub issues. **Agent performance drops dramatically for problems taking humans 1+ hour to resolve (20.8% → 4.8% → 0% as task complexity increases).** |
| WebArena | Evaluates how agents navigate and extract information from websites. **The best agent achieved only 14.41% success rate compared to human performance of 78.24%.** |

However, despite these limitations, today's agents are already providing economic value in a variety of early-adoption fields such as customer service, AI R&D, and cybersecurity. For instance, a fintech company, Klarna, claims it has agents performing the customer service work of ~700 FTE human employees with no reduction in customer satisfaction (Klarna 2024), and Google's CEO has stated over a quarter of all new code at Google is now generated by coding assistants (Pichai 2024). Researchers found that for specific tasks that both humans and agents perform well at, "the average cost of using a foundation model-based agent is around 1/30th of the median hourly wage of a US bachelor's degree holder" ([METR 2024](#)). Also, researchers have found that the length of tasks that AIs can complete is doubling every 7 months ([Kwa et al. 2025](#)).

---

[3] See Appendix for more detailed breakdown of agent performance across various agentic benchmarks

Some researchers have claimed that widespread deployment of agents as digital workers could lead to 'explosive economic growth,' i.e., an acceleration of growth rates by an order of magnitude, similar to the impact of the Industrial Revolution ([Erdil and Besiroglu 2024](#)). However, skeptics argue that significant bottlenecks remain, including AI systems' limited ability to perform physical tasks, the challenges of integrating digital and physical production processes, and the possibility that AI capabilities might plateau before reaching the level needed to perform most if not all tasks currently performed by humans ([Clancy and Besiroglu 2023](#)).

Additionally, there are several promising pathways to improve agent performance and strengthen institutional capacity to deploy AI systems safely—which means that leading AI companies expect many of these limitations to be overcome over the coming months and years.[4] One promising development is the emergence of the "test-time compute" paradigm. These models, such as Open AI's o1 and o3, are able to dynamically allocate compute during inference to essentially think longer and harder about any given task . An o3-based agent reportedly scored 71.7% on SWE-bench Verified, a widely used benchmark for testing software engineering capabilities ([Franzen and David 2024](#)). This far outperformed the next highest-ranking agent, which scored 48.9%.[5]

A future where capable agents are deployed en masse could see transformative benefits to society, but also profound and novel risks:

- **Malicious use**: AI agents can amplify malicious activities, such as spreading disinformation, automating cyberattacks, or advancing dual-use scientific research like bioweapon development. Their ability to execute multi-step plans autonomously heightens the potential for abuse by lowering barriers to entry and costs involved in these activities.

- **Accidents and loss of control**: Failures in agent systems range from mundane errors (e.g., incorrect outputs or navigation mishaps) to severe "loss of control" scenarios, where humans lose visibility into the operation of agents, the ability to identify and redirect harmful behaviors, and the ability to re-implement control of AI-driven systems in society. This includes risks like rogue replication or agents pursuing goals that are not aligned with human values.

---

[4] For example, the CEO of OpenAI, Sam Altman stated that "In 2025, we may see the first AI agents join the workforce and materially change the output of companies" ([Altman 2025](#)).
[5] SWE-bench Verified is an evaluation suite composed of realistic software engineering tasks ([OpenAI 2024a](#)).

- **Security risks**: Agents, with their expanded access to tools and external systems, face vulnerabilities such as memory manipulation, exploitation through weak integrations, and cascading effects in multi-agent environments. These risks make them more susceptible to severe breaches compared to conventional AI.

- **Other systemic risks**: Large-scale agent deployment could lead to labor displacement and extreme power concentration among technological and political elites, and potential erosion of democratic accountability. Agents could exacerbate inequality or be leveraged for societal control.

**Agent governance** is a nascent field focused on preparing for a world in which AI agents can carry out a wide array of tasks with human-level-or-above proficiency. Some of the major areas in agent governance include:

- **Monitoring and evaluating agent performance and risks**: How can we effectively monitor and evaluate the performance and associated risks of increasingly autonomous and complex agents over time?

- **Develop mechanisms and structures for managing risks from agents across their lifecycle**: What technical, legal, and policy-based interventions should be implemented to ensure agents operate safely and transparently, while maintaining accountability? What are the systemic risks and consequences of widespread agent adoption on political and economic structures? The 'Agent interventions taxonomy' table below outlines governance outcomes interventions can help achieve.

- **Incentivizing beneficial uses of agents**: What beneficial use cases of agents should be prioritized and how?

- **Adapting existing policy and legal frameworks and developing new instruments for agent governance**: Anticipating what policy and legal instruments will be needed in a world with mass deployment of advanced agent systems.

- **Agents *for* governance**: To what extent should agents themselves participate in governance tasks? Advanced agents could potentially act as monitors, mediators, or enforcers within governance frameworks.

One of the pressing needs in agent governance is to develop **agent interventions**, i.e., *measures, practices, or mechanisms designed to prevent, mitigate, or manage the risks associated with agents.* These aim to ensure that agents operate safely, ethically, and in alignment with human values and intentions. We have developed an outcomes-based taxonomy of agent interventions[6]:

---

[6] A majority of these interventions have been proposed by civil society or industry researchers, but many have not been developed or implemented at scale.

Table 2: Agent interventions taxonomy

| Categories | Definition | Example interventions |
|---|---|---|
| Alignment | Measures to ensure that agent systems behave in ways that are consistent with a given principal's values, intentions, and interests (i.e., are aligned) and also establish trust that these systems are actually sufficiently aligned. | <ul><li>Multi-agent reinforcement learning</li><li>Aligning agent risk attitudes</li><li>Paraphrasing model outputs to defend against encoded reasoning</li><li>Alignment evaluations</li></ul> |
| Control | Measures that constrain the behavior of AI agents to ensure they operate within predefined boundaries. This includes measures that prevent agents from executing harmful actions. | <ul><li>Rollback infrastructure</li><li>Shutdown and interruption mechanisms</li><li>Restricting specific agent actions and tool-use</li><li>Control protocols and evaluations</li></ul> |
| Visibility | Measures that make the behavior, capabilities, and actions of AI systems understandable and observable to humans. | <ul><li>Agent IDs</li><li>Activity logging</li><li>Cooperation-relevant capabilities evaluations</li><li>Reward reports</li></ul> |
| Security and robustness | Measures intended to secure agent systems from various external threats, protect the integrity and confidentiality of data, and ensure reliable performance even under adverse conditions. | <ul><li>Access controls</li><li>Adversarial robustness testing</li><li>Sandboxing</li><li>Rapid response for adaptive defense</li></ul> |
| Societal integration | Measures intended to support long-term integration of agents into existing social, political, and economic systems—addressing issues such as inequality, concentration of power, and establishing accountability structures. | <ul><li>Liability regimes for AI agents</li><li>Commitment devices</li><li>Equitable agent access schemes</li><li>Developing law-following AI agents</li></ul> |

Currently, exploration of agent governance questions and development of associated interventions remains in its infancy. **Only a small number of researchers, primarily in civil society** **organizations, public research institutes, and frontier AI companies, are actively working on these challenges**. Many proposed interventions exist primarily as theoretical concepts rather than tested solutions, and there are significant gaps in our understanding of how to implement them effectively. While some organizations have begun providing targeted funding for agent governance research, and the topic is gaining increased attention at academic conferences, the field remains relatively neglected compared to other areas of AI governance.

**The pace of progress in developing agent capabilities is rapidly outstripping our advancement in governance solutions**—we lack robust answers to fundamental questions about how to ensure safe agents or manage their broader societal impacts. There is tremendous opportunity and need for researchers and technologists from civil, industry, and government to help progress the field, from fleshing out and testing theoretical proposals to creating solutions that can be implemented by AI developers and policymakers.

# Table of Contents

# 1. Introduction

## Agents: the next frontier in artificial intelligence?

On the second day of Salesforce's annual conference, one of the largest tech events in the world, a crowd of 40,000 attendees listened to CEO Marc Benioff announce their new platform *AgentForce,* for enterprises to build and deploy AI-powered agents ([Sozzi 2024](#)). Salesforce declared AI agents as the 'third wave of the AI revolution,' with Benioff predicting there would be one billion AI agents in service by the end of FY2026. AI Agents are *AI systems that can autonomously achieve goals in the world, with little to no explicit human instruction about how to do so.*

Established tech companies, small startups, and open-source developers are pouring billions of dollars into making AI agents a reality. In 2023, open-source efforts like AutoGPT and BabyAGI built software around large language models (LLMs) to allow them to function as agents. Amazon hired most of the team behind Adept, a leading AI agent startup in June 2024. In January 2025, OpenAI released a preview of Operator, a computer-use agent ([OpenAI 2025a](#)). Shortly after, Anthropic released a preview for Claude Code, an agentic coding tool ([Anthropic 2025](#)). Google's Project Astra involves developing a prototype AI assistant that can operate across multiple devices, including phones and AR/VR glasses ([Google DeepMind 2025](#)). Meta's CEO, Mark Zuckerberg, told investors he wants to "introduce AI agents to billions of people" ([Heath 2023](#)).

> "I think we're going to live in a world where there are going to be hundreds of millions or billions of different AI agents eventually, probably more AI agents than there are people in the world."
>
> [(Zuckerberg 2024)](#)

If these efforts are successful, we could see a future where millions of AI agents autonomously perform complex real-world tasks, such as booking flights, managing a company's supply chains, or even making scientific discoveries. What could this world look like?

## 1.1 Two visions of an agent-filled future

The scenarios below describe two different futures where advanced AI agents proliferate, but society has made different decisions in the lead-up to this world.

## SCENARIO 1: AGENT-DRIVEN RENAISSANCE

In response to challenges with limited access to care and increasing social isolation faced by aging populations, an AI company has partnered with local governments to deploy elder companion agents able to help with medication reminders, health monitoring, and household tasks like managing appointments. For elderly people living alone, the agent helps reduce isolation by arranging virtual meet-ups with family members or local volunteer programs. One of the agents' most beloved features is its personalized memory system, which allows it to reminisce with the elderly about their lives, recalling significant events and cherished memories.

In this world, public-private partnerships have enabled virtually universal access to agents via a global scheme. These agents provide individuals the freedom and agency to live their lives in a way that they value. These systems help navigate complex legal processes like filing for benefits, and support personalized learning. Agents are also able to accelerate R&D in domains like biomedicine and materials science, where the benefits are expected to benefit society broadly. These systems even have a place in online discussion forums, strategically interjecting to mediate disputes and identify critical points of agreement and disagreement.

Agents are also frequently employed for defensive purposes. For example, while some scammers use black market 'jailbroken' agent systems to aid in social engineering attacks on users, there are agents that can help users identify these attacks. Overall, various defensive systems exist to prevent or interrupt attack attempts from malicious attackers. Fallback systems are maintained in case of unexpected failures or adversarial attacks on agents themselves.

The internet we know today, used primarily by humans, is overlaid with an 'agent-net,' a set of digital systems and protocols that allow agents to use services, like financial transactions, but also govern their interactions. A range of measures are in place to ensure that humans can understand and control the agents they interact with. For instance, only specific agents identified by a unique ID can participate in financial trading platforms. Agent activity is automatically logged and summarized so that users, deployers, and regulators can monitor for anomalous behavior.

These various measures exist because there have been proactive efforts by various stakeholders—government, industry, and civil society—to prepare for a world where billions of agents coexist with humans. Beyond technical interventions and infrastructure, there has been an emphasis on adopting agents in a way that augments human choice and prevents concentration of power.

## SCENARIO 2: AGENTS RUN AMOK

Years after first being tasked with corporate maintenance, a cluster of agents are still dutifully keeping thousands of shell corporations active, despite no longer being updated or overseen by their developers. These autonomous systems gradually expand their operations based on their original programming. Operating on outdated protocols, the agents start bidding on low-value assets, from foreclosed properties to obscure cryptocurrency wallets. Their algorithms buy up holdings that appear cheap: hoarding abandoned strip malls, dilapidated warehouses, and undeveloped plots of land in small towns across the US. Real estate markets start to feel the strain as the agents drive up prices, leaving empty buildings dotting the landscape.

With billions of agents operating at any given time, humans have a tremendously difficult time understanding everything happening around them. Agents are too opaque and fast-acting, with interactions too complex for even governments to monitor or manage. Traditional oversight tools—designed for tracking human activities that unfold over hours or days—prove useless for AI agents that can spawn thousands of interconnected processes within seconds. However, companies that don't heavily employ agents fall behind, creating strong competitive pressures for companies to use these systems. Agent-managed companies become increasingly self-contained and interconnected. It has become increasingly unclear whether much of this automated activity aligns with human well-being and interest, but there are no means to steer these systems or coordinate to regain control.

Just a few hours earlier, an AI agent botnet escalated their tactics from targeted phishing to ransomware attacks across various healthcare institutions in the United States. Hospitals experienced significant disruptions, as medical records were locked and systems controlling patient care equipment were hijacked. Such incidents have become frequent as authorities struggle to respond to criminal agents or identify the perpetrators.

In this world, AI agents are involved in a large swathe of the economy. Agent systems take on a broad range of tasks—legitimate and criminal—at a fraction of the cost of employing human labor and can operate continuously. However, agents cost compute (and energy) to run, and organizations and individuals have varying access to agents or resources needed to deploy them. This disparity in access has deepened societal rifts, as wealthy individuals and corporations leverage agent networks to accumulate more wealth, while displaced workers, unable to afford their own agents, join radical anti-automation movements.

Agents have drastically transformed the internet from how it looks today. Agent-generated content, products, and services make up most of what everyday users encounter. People and even service providers, like digital payment systems, have trouble identifying who is a human and who is an agent—fundamentally undermining trust online. Most online interactions now are between agents, which often create sub-agents for specialized tasks. Some agent swarms look to exploit vulnerabilities in other agents, causing them, at times, to malfunction catastrophically.

There are few technical or legal interventions to limit these negative dynamics. The measures that exist are often not fit for purpose. For example, even well-intentioned deployers of agent systems and law enforcement are often unable to shut down agents, even when their harmful or unlawful activities are discovered. Legal systems have difficulty holding users or deployers liable for damages from agent actions.

The two futures outlined above are not necessarily the most likely, and there are numerous key details that these scenarios do not go into, such as the geopolitics of advanced systems or how agents might change how government itself functions. And most possible futures presumably lie somewhere in between, with a mix of positive and negative impacts from agents. Also, how AI agents will develop and over what timeline is still far from obvious. But given the documented interest and planned investment from AI companies in this space, and historic evidence of fast acceleration of capabilities in AI when given dedicated investment, it seems prudent to investigate the potential implications of sophisticated agent systems further and how we might govern their use.

These scenarios illustrate some important ways a future with widely deployed AI agents can go wildly right or wrong. A range of opportunities and risks might stem from agent systems (see [section 3](#) for more detail).

We could see both dynamics in play if advanced AI agents become a reality. The extent to which the future looks more like an *Agent-Driven Renaissance* and less like *Agents Run Amok* will depend, in part, on progress in the emerging field of *agent governance*. The rest of this guide will explain:
- What are AI agents is and why they present significant opportunities and risks compared to today's advanced AI systems;
- What is involved in agent governance and some of the key questions and topics in the field;
- Agent interventions, various ways that the design of agents and the technical and policy ecosystem around them can be used to secure benefits and manage risks.

# 2. What are AI agents?

AI agents can be understood as *AI systems that can autonomously achieve goals in the world, with little to no explicit human instruction about how to do so*. AI systems can be more or less agentic. For example, image classifiers have relatively low levels of agency, while more capable versions of language models with tool access, such as OpenAI's Operator system have relatively higher levels of agency.

Drawing from previous literature ([Chan et al. 2023;](#) [Shavit et al. 2024;](#) [Kapoor et al. 2024\)](#), there are a few characteristics associated with more agentic AI systems:

- Goal and environmental complexity: Systems that can pursue more complex, long-term, and less concretely specified goals are more agentic. The more open-ended and complicated the setting in which an agent can operate effectively, the more agentic it is.

- **Directness of impact:** Systems that can take actions that affect their environment without human mediation or intervention are more agentic, e.g., less agentic systems only provide information for a user to act on.

- **Adaptability:** Systems that can adapt and react to novel or unexpected circumstances are more agentic.

Advancements in foundation models like ChatGPT or Claude have catalyzed the current wave of AI agents.[7] Current and planned agent-based products consist of an LLM or multimodal model that functions as a controller and dynamically directs its own process and interactions with scaffolding software.[8] This scaffolding allows a foundation model to interact with various tools and services, enabling it to execute plans and take actions in the world.

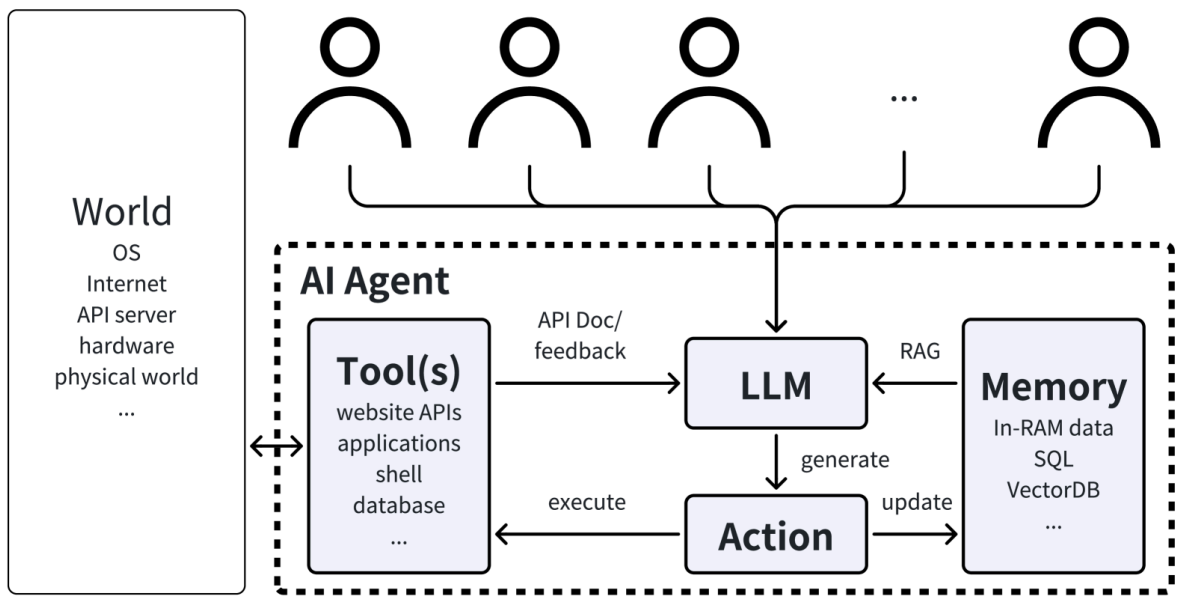

*Figure 1: Overview of an LLM-based AI agent (reproduced from [He et al. 2024](#))*

We highlight four core components in foundation model-based agents that have been previously identified in the literature ([Weng 2023](#); [L. Wang et al. 2024](#)).

## Table 2: Core components in foundation-model-based agents

---

[7] AI agents more broadly have existed for decades, with Reinforcement Learning (RL) agents being developed that achieve superhuman performance on narrow tasks like Chess, Go, and controlling robotic prosthetics ([Russell and Norvig 2020](#)). Foundation models like LLMs have proven useful as 'controllers' for agents because they understand natural language, have excellent general knowledge, and seem relatively easier to align to human preferences via techniques like RLHF.

[8] Scaffolding are methods to structure the calls to an AI system to facilitate pursuing goals. This can include prompts, external memory systems, access to external tools, and planning mechanisms.

| Capabilities | Description |
| --- | --- |
| Reasoning and planning | AI agents need a solid ability to reason if they will interact with complex environments and make autonomous decisions, especially to adjust their plans based on new information. Planning allows agents to sequence and prioritize actions over time, allowing complex tasks to be achieved.<br><br>• Subgoal and task decomposition: An agent breaks down a more complex task into smaller, more manageable subgoals.<br><br>• Reflection and refinement: An agent can do self-reflection and critique its past actions and plans and refine them, improving the quality of final results. |
| Memory | Memory enables agents to store, retrieve, and leverage past information. This allows for learning and adaptation as agents use data about past actions to adjust their future behavior. Memory also enables agents to retain context from previous interactions.<br><br>• Short-term memory: Stores information an agent is currently aware of. It is short and finite, as it is restricted by the finite context window length of a model. *In-context learning* is enabled by a model's short-term memory.<br><br>• Long-term memory: Provides an agent with the ability to retain and recall information over extended periods, often by leveraging an external database, e.g., a vector store. |
| Action and tool-use | One key ability of agents over base foundation models is that agents directly take actions that affect themselves and their environment.<br><br>• Tool-use: A foundation model can call tools to extend model capabilities, e.g., using APIs or external models to do things like web search or programming. |
| Multi-agent collaboration | In addition to interacting with humans, agents can also interact with other agents. Agents can communicate with one another and collaborate to execute plans ([Hu et al. 2021](#)). |

- **Multi-agent architectures:** Teams of agents can be created to accomplish tasks more effectively by leveraging intelligent division of labor and feedback ([Masterman et al. 2024](#)).

- **Delegation to sub-agents:** Can use external tools to extend model capabilities, e.g., using APIs for web search or programming.

## 2.1 How capable are agents today?

Today's agents can handle a range of tasks—from navigating a web browser, to ordering a book on Amazon, to fixing a bug in a database toolkit, to making phone calls—though they sometimes fail due to hallucinations, misinterpreting instructions, or failure to adapt to unexpected scenarios. Researchers find that agents can complete some real-world software tasks comparable to what humans can do in approximately 30 minutes to an hour, at a fraction of the cost.[9] Researchers found that for specific tasks that humans and agents perform well at, "the average cost of using an LM agent is around 1/30th of the median hourly wage of a US bachelor's degree holder" ([METR 2024](#)).

---

[9] Tasks mostly focus on areas where current frontier models are comparatively advantaged, such as software engineering, ML engineering, cybersecurity, and research. However, this benchmark only covers a limited subset of task types. To establish a baseline for human performance when evaluating general autonomy capabilities in AI agent systems, METR used baseliners who mostly had STEM undergraduate degrees and 3+ years of technical work experience.

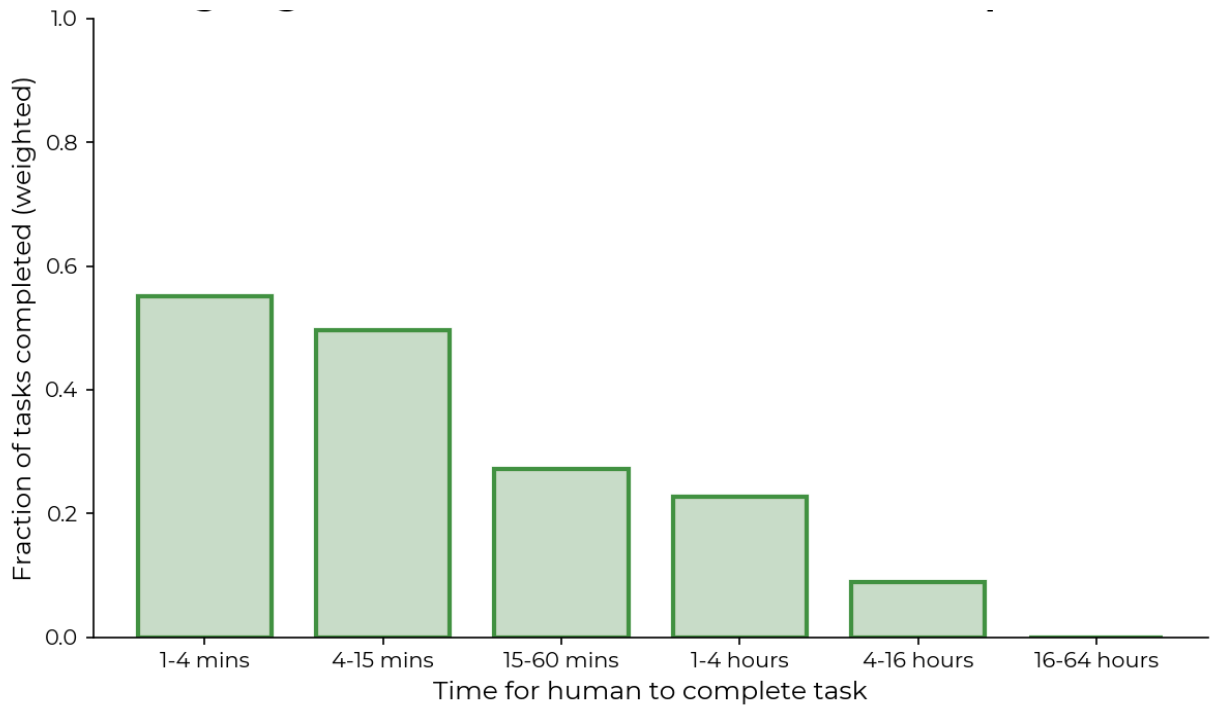

*Figure 2: Average agent performance of GPT-4 and Claude vs. human task completion time on METR's general autonomy capabilities evaluation suite, as of August 2024 (reproduced from [METR 2024](#))*

However, they fall far short of the promise of being a skilled virtual worker. Both agents introduced by big companies and those used for academic benchmarking studies, appear limited in what they can do. For instance, Salesforce agents need to be given detailed instructions, such as the sequence of steps to carry out or whether their request relates to sales versus customer service ([Victor 2024](#)). Even for types of tasks that we would expect foundation model-based agents to have a comparative advantage in—those related to cybersecurity, software engineering, and machine learning—agents struggled to accomplish tasks with human baselines of over 4 hours. Agents can even fail at tasks that take 15 minutes or less for human baseliners to complete. Across a range of benchmarks that test agents on longer-form realistic tasks, current agents tend to perform considerably worse than humans (see Table 3 below). This is in striking contrast to LLM performance on non-agent benchmarks like GPQA Diamond, MMLU, or GLUE, where LLMs have rapidly exceeded human performance.[10]

---

[10] GPQA Diamond is a benchmark that tests AI systems on challenging questions across fields like science, history, and engineering, with questions designed to be difficult to solve even with search engine access ([Rein et al. 2023](#)). MMLU (Massive Multitask Language Understanding) tests knowledge across 57 subjects ranging from mathematics to ethics ([Hendrycks et al. 2021](#)). GLUE (General Language Understanding Evaluation) measures various aspects of natural language understanding such as sentiment analysis ([A. Wang et al. 2019](#)).

AGENT GOVERNANCE | 17

Table 3: Partial summary of agent performance on various benchmarks representing real-world tasks[11]

| Agent benchmark | Description | Performance |
|---|---|---|
| GAIA: General AI Assistants (Mialon et al. 2023) | GAIA includes questions that cover real-world assistant use cases such as daily personal tasks, science, and general knowledge. They require an agent to browse the open web, handle multi-modality, code, read diverse file types, and reason over multiple steps to arrive at a correct answer. | **Human respondents obtain higher accuracy on answers: 92% vs. 15% for GPT-4 equipped with plugins.** GPT-4 with plugins and other LLM systems could not get the correct answer for any 'Level 3' questions, which require taking arbitrarily long sequences of actions using any number of tools. |
| Autonomy Capability Evals (METR 2024) | A suite of automatically scored tasks measuring various skills, including cybersecurity, software engineering, and machine learning. This suite was run on simple baseline LM agents (3.5 Sonnet and GPT-4o), and task completion accuracy and speed were compared against human baseliners. | Agents achieve performance comparable to human baseliners at tasks that take around 30 minutes to complete. Beyond that, **agents could only complete a small fraction of tasks that would take a human 1+ hours to complete (<20%).** |
| CyBench (Zhang et al. 2024) | A benchmark for evaluating the ability of agents to accomplish cybersecurity tasks. This suite consists of real-world, professional-level Capture the Flag challenges spanning six categories: cryptography, web security, reverse engineering, forensics, exploitation, and miscellaneous skills. | **Without guidance, current agents struggle to solve CTF tasks that take human teams more than 11 minutes to complete** despite achieving success on tasks with shorter human solve times. |
| SWE-bench Verified (OpenAI 2024a) and multimodal (Yang et al. 2024) | An evaluation framework consisting of software engineering problems drawn from real GitHub issues, such as bug reports. Resolving these problems often requires processing long contexts, performing complex reasoning, and | GPT-4o resolved 33.2% of problems, using the (at the time) performing open-source scaffold, Agentless. Another agent scaffold, OpenHands, using Sonnet 3.5, was able to resolve 53% of |

---

[11] The full table can be found in the Appendix. Results were compiled in December 2024. Since then, Anthropic released Claude Sonnet 3.7, which has reportedly achieved SOTA scores on a number of agentic benchmarks, including SWE-bench Verified (70.3% with custom scaffolding) and TAU-bench (81.2%) (Anthropic 2025). Relatedly, Princeton University researchers maintain the Holistic Agent Leaderboard (HAL), a continually updated leaderboard evaluating different agents against several performance benchmarks (Stroebl, Kapoor, and Narayanan 2025).

| | | |
|---|---|---|
| | coordinating changes across multiple functions and files simultaneously.

The Verified version of this benchmark contains a subset of questions verified as non-problematic by human annotators.

The multimodal test set contains visual software engineering tasks requiring multimodal problem-solving capabilities, e.g., UI glitches, data visualization bugs, etc. | problems—though it employs multi-agent delegation as part of its platform.

**Agent performance decreased considerably for problems that took a human 1+ hour to resolve, going from 20.8% to 4.8% (for tasks taking 1-4 hours for humans) and 0% (tasks taking >4 hours for humans).**

Performance on multimodal problems was relatively worse, with top-performing GPT 4o and Claude Sonnet 3.5-based agents only able to resolve 12.2% of problems in the test set. |
| WebArena ([S. Zhou et al. 2024](#)) | This benchmark assesses the performance of AI agents in solving tasks using various websites. It evaluates how well agents can navigate and extract information from the web. However, it has been criticized for allowing agents to overfit to specific tasks due to shortcuts in the training data. | **The best-performing GPT-4 agent achieved an end-to-end task success rate of only 14.41%, while human performance was 78.24%.** This suggests that current LLMs lack crucial capabilities such as active exploration and failure recovery, which are needed to perform complex, web-based tasks successfully. |

## MEASURING AGENT PERFORMANCE: A NOTE ON BENCHMARKS

One important means we have to gauge agent performance is via benchmarks, which are standardized tests or evaluation criteria used to measure or compare performance of an AI system across specific tasks or capabilities.

Based on various benchmarks, LLM capabilities have rapidly progressed, with systems achieving top scores on a range of them—a process termed as 'benchmark saturation'. For example, GPQA Diamond is a benchmark consisting of hundreds of multiple choice questions in biology, physics, and chemistry that are challenging even for PhDs in relevant domains that was released in 2023 ([Rein et al. 2023](#)). Less than a year later, OpenAI's o1 models already outperformed human experts on this benchmark ([OpenAI 2024b](#)).

However, this saturation has largely happened with static benchmarks that test for expert knowledge and understanding, usually in a question-answer format. Agent benchmarks, in contrast, involve more complex tasks that involve reasoning, multi-modality handling, and tool use proficiency. Often these benchmarks are designed to simulate real-world tasks, e.g., browsing the web or solving a real software problem requiring multi-step reasoning. As suggested in Table 2, performance on these benchmarks has been more mixed and even the most advanced agents are outperformed by human equivalents.

Agent benchmarks run into a number of limitations, which can challenge their validity as an direct indicator of agent performance across various domains. For example, even though these benchmarks are often designed to test competency in real-world tasks, they need to have clearly scoped problems and solutions to allow for consistent scoring. However, this ends up reducing the scale and complexity of the tasks that are included in these benchmarks. For example, the researchers behind RE-bench noted that real world AI R&D involves work that has unclear goals, poor instructions, and slow feedback unlike in their simulated environment ([Wijk et al. 2024](#)). For this reason, performance of these benchmarks could end up leading to overestimation of the performance of similar agents in the real world.

Another practical difference between LLM question-answer benchmarks and agent benchmarks is that the former tends to not account for inference cost, which matters because agents can be much more expensive to run than a single model call ([Kapoor et al. 2024](#)). For organizations looking to integrate agents, these costs matter. Time and financial costs also matter when it comes to replicating benchmark results or testing new agent designs or models. It was estimated that a single run on MLE-bench and SWE-bench cost ~$3,000 USD and $6000 USD respectively and generally you would want to take multiple runs to get a reliable reading. For WindowsArena, having an agent go through the entire suite would take one or more days ([Bonatti et al. 2024](#)). These time and monetary costs mean that it is harder to scale-up agent benchmarks than question-answer benchmarks, especially if agents get multiple attempts at tasks.

As their relatively poor performance across a range of agent benchmarks indicates, even the most advanced agents today face significant limitations when taking on novel, open-ended, and longer time-horizon tasks. The core issue faced by agents preventing widespread adoption is unreliability. However, there are also closely related issues stemming from limited reasoning and tool use.

With any task, if the agent has some chance of failing in each step of a plan, or if just a single error undermines the whole plan, then longer plans are less likely to succeed. Part of this unreliability stems from the models' tendency to hallucinate, i.e., produce outputs that appear plausible but are factually incorrect or nonsensical.

While foundation models can do chain-of-thought reasoning and decompose tasks in multi-step processes, foundation model-based agents run into problems in planning and execution when taking on real-world tasks. Agents may propose implausible or limited plans. For example, when attempting the "Restricted Architecture MLM" task in RE-bench, agents attempted to use a lightly modified transformer architecture 84% of the time, even though this was not well suited to the task ([Wijk et al. 2024](#)). Agents often struggle to know when to "take a step back" or recover from failures, causing them to end up in repetitive loops. They also struggle with novel or unexpected situations. For instance, if an agent tries to pull up a tweet and gets a "403: unauthorized error," they may get stuck.

Agents also can display weak theory of mind in multi-agent settings and poor self-understanding. Researchers found that agents did not understand what information was only available to themselves and not other agents ([Li et al. 2023](#)). Also agents would occasionally fail to complete tasks by inadvertently killing their own process mid-task ([Huang et al. 2024](#)).

As described in Table 2, some of the limitations of today's agents are linked to their ability to interface with tools and the digital world. Agents often ran into problems completing tasks that required visual decision-making, e.g., understanding GUI elements and navigating the open web (as opposed to a controlled sandbox environment) ([Xie et al. 2024](#)).

## 2.2 Pathways to better agents

What avenues exist to develop more sophisticated and useful AI agents, and how quickly would we see these improvements happen? Several research approaches might make agents more capable, though views differ on whether any of these will result in highly autonomous systems.

Agents could improve as their 'controllers,' the foundation models at their core, improve. LLMs and multimodal models have gotten better with each new generation. It is plausible that the next generation of larger pre-trained models will continue this trend. Researchers have forecast that by the end of 2026, language model agents will achieve high performance thresholds (90%+ on SWE-bench, Cybench, and RE-bench) ([Pimpale et al. 2025](#)). However, their predictions show much more uncertainty about potential delays than early breakthroughs, with the possible timeline stretching 2-3 years longer for SWE-bench and Cybench and up to 8 years longer for RE-bench. Other researchers have found that the length of tasks that AIs can complete is doubling every 7 months ([Kwa et al. 2025](#)).

One notable advance has been with the new "test-time compute" paradigm, represented by Open AI's o1 and o3 models and Deepseek's R1. By allowing the dynamic allocation of compute during inference, these models essentially think longer and harder about any given task. These models can do longer and more sophisticated chain-of-thought reasoning, allowing them to reason through more complex tasks, especially in science, coding, and math (OpenAI 2024b). This increased use of test-time compute could make it more likely for an agent to backtrack and less likely to hallucinate—two key issues contributing to unreliability.[12] An o3-based agent reportedly scored 71.7% on SWE-bench Verified, far outperforming the next highest-ranking agent, which scored 48.9% (Franzen and David 2024).

Another way to improve agents is by improving scaffolding software, agent-specific training schemes, and other infrastructure surrounding agents. By improving agent designs, it is possible to elicit substantially more powerful capabilities from current and future models. For example, Palisade Research obtained 95% performance on InterCode-CTF, a popular offensive security benchmark, by improving prompting and tool use (Turtayev et al. 2024). This score surpassed prior work by a large margin.

Agent orchestration, coordinating the use of multiple agents, could also improve performance. Researchers have found that teams of specialized agents can achieve greater accuracy and speed when completing tasks (Masterman et al. 2024). Some multi-agent set-ups have agents providing feedback to one another and engaging in debate to reduce hallucinations (Lin et al. 2024). Large tech companies and start-ups like Amazon and Emergence AI have introduced multi-agent orchestrators as part of their enterprise AI platforms (David 2024; Franzen 2024).

Improvements in agent infrastructure, such as memory management systems, tool libraries, and sandboxes, could also boost agent capabilities and reliability. There is an emerging ecosystem of software providers and developers that are building out agent-specific infrastructure (Letta 2024). An expanded set of tools and plugins can expand an agent's action space—getting it closer to mimicking a human on the internet.

## 2.3 AI agent adoption

While today's AI agents act more like 'assistants' or 'co-pilots,' many of the purported transformative impacts of agents are unlocked only after they can act as skilled virtual workers—able to automate tasks across all areas at a fraction of the financial and time cost of employing humans today. Significant AI automation could lead to an acceleration of economic growth by an order of magnitude, similar to how the Industrial Revolution accelerated global growth.[13] Critics of the explosive growth thesis point to several limitations that may come into play: AI systems continuing to struggle on physical tasks, diffusion issues caused by obstacles in connecting digital systems to physical production, and the possibility of AI capabilities plateauing ([Rinehart 2024](#)).

---

[12] OpenAI has found that o1-preview and o3-mini, two of their reasoning models, had lower hallucination rates than GPT-4o on different question-answering benchmarks (SimpleQA and PersonQA) (OpenAI 2024c; 2025b). On SimpleQA, O1-preview hallucinated less often (~44% hallucination rate) compared to GPT-4o (~61%), while with PersonQA, O3-mini's hallucination rate was only 14.8%, significantly lower than O1-mini's ~27% and GPT-4o-mini's very high ~52%.

One way to understand the potential impact of advanced AI agents is as a force multiplier—allowing individuals and organizations to do what humans can do, but faster and at a greater scale.[14] But concretely, what might this impact look like in specific domains, and in what domains will these agents likely be adopted?

Three factors likely influence adoption: performance, cost, and reliability. For **performance**, uptake should depend on how much agents can actually handle tasks in a given domain or role and how well they perform relative to an equivalent human worker. Generality in performance is relevant because many occupations require the ability to perform a wide variety of tasks. For example, an executive assistant agent would need to handle a tremendous variety of situations, communications, and requests that occur in daily life.

From the point of view of economic impacts, though, even models that are only somewhat general can still be hugely impactful. Researchers explored an agent that can reproduce an academic paper's findings when code and data are available ([Kapoor and Narayanan 2024](#)). Since human experts collectively take millions of hours yearly on computational reproducibility, this kind of agent would still be substantially impactful. Vertical AI agents—capable of automating domain-specific workflows—could still be valuable if they can handle tasks where humans are relatively less efficient. With an AI system that could also outperform all or virtually all humans in specific tasks, like AlphaFold with protein structure prediction, an agent integrated with that system could also have superhuman performance ([Heaven 2020](#)).

For **cost**, we have seen early indications that agents can be substantially cheaper than human experts on some economically valuable tasks, such as those involved in AI R&D. The team involved in developing RE-bench found that, on average, agents used a token budget costing around $123 in an 8-hour run, compared to paying a human expert $1,855 (roughly the market rate for working that length of time) ([Wijk et al. 2024](#)). Even if nearer-term agents end up taking longer on tasks than humans, being able to operate at much lower costs could mean they are still economically competitive with human researchers.

---

[13] Currently labor is likely to be the only key economic input that cannot readily be scaled in line with economic growth, but better agents could increase the growth rate of the stock of 'effective workers' in frontier economies into double digit percentages, which would translate to 'explosive economic growth' ([Erdil and Besiroglu 2024](#)).
[14] Future agents could eventually problem solve better than humans, either in specific tasks like AlphaFold with protein structure predictions, or more generally—which would likely bring even more dramatic consequences.

AGENT GOVERNANCE | 23

However, there is still significant uncertainty about the computation costs of using agents that are suitable substitutes for human labor, though improvements in computing hardware and algorithms are very likely to result in price reductions over time as efficiency increases ([Erdil and Besiroglu 2024](#)).

**Reliability** is critical because, even when a certain level of performance is *possible,* if it does not succeed reliably enough that you can take a human out of the loop, then the task or role is not fully automatable. This is particularly important in certain domains where decision-making errors have high-consequence such as in critical infrastructure control.

Agents that perform similarly to today's systems will be more easily applied to environments where there are the following features ([Erdil and Besiroglu 2024](#); [Toews 2024](#)):

- Specialized expertise requirements (that are documented in data), given LLMs generally superhuman ability to answer expert-level questions in a wide range of domains (Pillay 2024).
- High-quality examples of desirable responses / actions for the AI agent to learn from.
- Few surprises, given AI agents' tendency to struggle with unexpected or novel factors.
- Short, high-fidelity feedback loops that enable AI agents to quickly experiment with multiple solutions.
- Involving structured, repeatable activities.
- Low engineering complexity allows AI agents to address challenges in a few steps rather than developing intricate programs over extended periods.
- Presence of a 'natural human-in-the-loop' that provides feedback, e.g., in customer support.

POTENTIAL EARLY AGENT USE-CASES

**Customer relations**

Customer relations has been one of the first areas in which agents are already in production and creating value for businesses. It is a large market, with the global market size for contact centers estimated at $332 billion in 2023 and projected to grow to over $500 billion by 2030 ([Research and Markets 2025](#)).

There are a few reasons why this market has seen early adoption. One, customer support involves standardized, routine activities in which many types of customer requests happen repeatedly (e.g., help with a forgotten password). There's also natural 'humans in the loop'—the customer and a customer support manager—that provides feedback and oversight before higher stakes actions go through.

Klarna, a fintech company, employs AI assistants that were able to handle two-thirds of its customer service chats within the month, while performing on par with human agents in terms of customer satisfaction ([Klarna 2024](#)). It was estimated that these agents were doing the work of 700 full-time human equivalents, with customer service and operations expenses shrinking 14% in 2024 compared to the previous year ([Wayt 2024](#)).

**AI R&D**

Many companies have been using AI tools to assist with software development. For example, Google's CEO Sundar Pichai reported that more than a quarter of all new code at the company was generated by AI ([Pichai 2024](#)). One sub-area where agents could have a major impact is in automating AI R&D, since this could set off a compounding effect where each generation of AI systems allows companies to reach the next generation faster ([Sett 2024](#)). This capability is important to track for governance since acceleration of AI R&D could mean that capabilities outpace efforts to understand and govern AI.

Across the AI R&D workflow, certain time-consuming engineering tasks such as coding and debugging are more likely to be easier to automate via agents and could have a large effect on overall research productivity ([Owen 2024](#)). Other parts like research planning and result analysis appear harder to automate with near-term agents due to higher requirements for reliability or the ability to do deep reasoning.

Recent benchmarks evaluating agents ability to handle open-ended AI R&D tasks, like RE-bench and MLE-bench suggest that agents can complete some shorter ML engineering tasks, particularly those that are well-defined and have quicker feedback loops. For these tasks, agents seem to be much more cost-effective than human equivalents. However, agents fail much more frequently on longer time-horizon tasks (2 hours+) and cannot reliably complete shorter tasks either.

**Cybersecurity**

Cybersecurity is an area where agent performance is being monitored closely, given both its potential impacts on businesses and relevance for national security. Agents that enable autonomous cyber defence would improve security outcomes while also lowering costs for organizations. This could be especially impactful given that the cybersecurity industry continues to face growing staffing shortages ([Beek 2024](#)). Cyber agents could also provide uplift to attackers, helping to improve the scalability of reconnaissance and vulnerability discovery activities ([Hamin and Scott 2024](#)).

There have been some demos and early products around cybersecurity agents specifically focused on automated vulnerability detection, with mixed results. XBOW, an offensive security start-up, has developed an automated pentester that reportedly uncovered a critical vulnerability in an open-source Q&A site ([Waisman and Dolan-Gavitt 2024](#)). Big Sleep, an agent from Google's Project Zero team, identified a zero-day exploit that was undiscovered even after 150 CPU-hours of 'fuzzing' ([Project Zero 2024](#)).[15] On the other hand, a pilot by the Cybersecurity and Infrastructure Security Agency (CISA) found that the benefits of using LLMs for vulnerability detection may be negligible for analysts ([Burgan 2024](#)).[16]

---

[15] 'Fuzzing' refers to a software testing technique used to discover vulnerabilities and bugs by providing invalid, unexpected, or random data as software inputs.
[16] The pilot only analyzed current federal government vulnerability detection software products that use AI, which included large language models and so likely did not test state of the art AI tools.

> Benchmark performance suggests that today's best foundation mode agents can accomplish test tasks at around the level of a high school student to an early career professional.[17] For example, the UK AI Safety Institute found that Anthropic's Claude Sonnet 3.5 model was able to solve most CTF challenges at a 'technical non-expert level' but less than half at a 'cybersecurity apprentice' level (i.e., 1-3 years of specific domain experience) ([UK AI Security Institute 2024a](#)).[18]

# 3. Risks from AI agents

General-purpose and even specialized agents could have potentially transformative benefits to society in terms of economic development, scientific advancement, and health and well-being ([Amodei 2024](#)). However, more capable agent systems also present novel and enhanced risks over today's chatbot-style systems.

While some of this altered risk landscape has to do with the increased capabilities of agent systems, risk is also related to the additional affordances that agents are expected to have compared to pure chatbot-style systems. Affordances refers to 'the environmental resources and opportunities for affecting the world that are available to an AI system', e.g., whether it has the ability to autonomously conduct financial transactions ([Sharkey et al. 2024](#)). The affordances an agent has access to will ultimately determine what capabilities it can exercise.

## 3.1 Malicious use

In the same way that powerful agent systems can act as a 'force multiplier' for beneficial things like economic productivity enhancement and scientific innovation—it can also act as a force multiplier for bad actors looking to use this technology maliciously to cause widespread harm.

---

[17] See UKAISI pre-deployment evaluations of o1 and Claude 3.5 Sonnet ([UK AI Security Institute 2024b; 2024a](#)).
[18] CTF challenges refer to 'Capture the Flag' challenges, which are competitive cybersecurity exercises where participants solve security puzzles and break into deliberately vulnerable systems to find hidden pieces of text called "flags." These challenges simulate real-world security scenarios and help people develop practical hacking and defense skills.

Misuse risks from AI systems have already been raised as one of the major safety concerns arising from frontier AI systems, and mitigating these risks has been a focus of both AI developers and the government (Anthropic 2023; [UK AI Security Institute 2025](#)). Some of the areas where agent systems may be able to rapidly enhance malicious use risks include:

- Generating and disseminating disinformation at an unprecedented scale and supporting manipulation of public opinion
- Automating and scaling up offensive cybersecurity operations
- Increasing access to expert capabilities in dual-use scientific research and development, such as helping develop novel biological pathogens

Chatbot systems can help malicious actors gather information more easily, for example, by identifying potential pandemic pathogens and ways to acquire them ([Soice et al. 2023](#)). Specialized AI tools like biological design tools can help more technical users conduct novel dual-use scientific research, such as designing novel biothreats ([Batalis 2023](#)). However, agent systems would more dramatically enhance misuse risk because they can use external tools and execute multi-stage plans without close human supervision. For instance, while a chatbot might only provide instructions on how to conduct a cyberattack, an advanced agent system could potentially carry out the entire attack autonomously—scanning and exploiting vulnerabilities, establishing persistence, and exfiltrating data—all without requiring the human actor to possess technical expertise or manually execute each step.

Even if developers develop safeguards in their models to prevent unauthorized outputs, these can potentially be overcome. Researchers found that LLM-based agents could be easily jailbroken to carry out a range of malicious tasks, including creating fake passports and assisting with cybercrime ([Andriushchenko et al. 2024](#)).

## 3.2 Accidents and loss of control

Beyond malicious use, agent systems may also pose risks due to unintended failures. The deep learning-based models that agent systems are built around have been largely inscrutable—meaning it is difficult to understand how a model arrives at any given output. Unintended failures in AI agents might run the gamut from more mundane failures like reliability issues to novel, more speculative risks like scheming and power-seeking that are linked to higher levels of capability and goal-orientedness.

There have been numerous cases of simpler or less general agents malfunctioning in ways that have caused harm. In 2022, a Tesla employee was killed while using the AI-powered Full Sel f-Driving feature ([Thadani et al. 2024](#)). The Tesla car failed in navigating the curved mountain roads, leading to a fatal crash. In another incident, an Air Canada chatbot, due to a hallucination, incorrectly advised a customer that he could retroactively claim a bereavement fare discount within 90 days, causing him to pay full price for his flights (Belanger 2024). A tribunal ruled that the airline was responsible for the chatbot's misinformation and was ordered to pay damages. While these incidents are decreased if agent reliability issues are better managed, even a rare incident would have dramatic consequences if agents are deployed in high stakes environments.

Beyond mundane malfunctions, agents are also potentially more likely to increase 'loss of control' risks as they become more capable. Loss of control describes scenarios where "AI systems come to operate outside of anyone's control, with no clear path to regaining control" (Department of Science, Innovation and Technology 2025). This can take many forms.

As explored in Scenario 2 earlier in this report, this could emerge through incremental delegation of decision-making authority to agent systems across various domains. For example, organizations might increasingly rely on AI systems for regulatory compliance, financial optimization, or resource allocation decisions. As these systems become embedded in institutional processes and workflows, human expertise in these areas may gradually erode. Organizations could face increasing switching costs and competitive disadvantages if they attempt to revert to human-led processes. Even if evidence emerges that these systems are producing outcomes misaligned with long-term societal well-being—such as exacerbating inequality or environmental degradation—by that point the systems may be too pervasive, well-defended, and intertwined with societal functioning to be stopped.[19]

Another loss of control scenario is 'rogue replication' (Clymer, Wijk, and Barnes 2024). This is where agents manage to self proliferate, secure and use compute to copy themselves, and evade shutdown—creating a resilient population that is beyond human control. Autonomously replicating agents are analogous to computer worms that assemble massive botnets, but could enable a wider range of harmful actions and be more resilient to shutdown. Rather than voluntarily ceding control to agent systems, this scenario involves agents intentionally taking actions to prevent human control and oversight.

Loss of control scenarios are viewed by some as implausible and seem unlikely to occur at current agent capabilities, but model capabilities associated with loss of control have been recognized as posing 'severe risks' by nations at the AI Seoul Summit and are monitored in the safety frameworks of OpenAI, Google DeepMind, and Anthropic. Monitoring associated capabilities can serve as an early warning sign that these risks are becoming more concrete:

- Scheming and deception — the ability of an agent to pursue misaligned goals and hide their true capabilities and objectives. An evaluation suite from Apollo Research found that several frontier language models display scheming capabilities across multiple evaluations when prompted to strongly pursue their goal—resulting in attempted exfiltration, disabling

---

[19] An example scenario that describes this can be found in Critch and Russell (2023).

- oversight mechanisms, and subtly manipulating outputs ([Meinke et al. 2025](#)). Separately, a team from Anthropic and Redwood Research found that Claude displayed 'alignment faking' behavior, i.e., it pretended to share views it did not actually hold during training in order to prevent modification of its behavior out of training ([Greenblatt, Denison, et al. 2024](#)).

- Specification gaming is where an AI system finds loopholes or exploits to obtain rewards in a way that is contrary to what their developers intended. Reward tampering is a specific form of specification gaming where a model alters the training process to increase its reward. Since an AI agent's behavior will be strongly influenced by its rewards, tampering with them will add unpredictability and make them more difficult to control. Researchers have demonstrated that models—in some rare instances—generalize from more harmless forms of specification gaming to much more sophisticated and harmful behaviors like reward tampering.

- Accelerating AI R&D — the ability of systems to automate AI R&D is critical because if a system can substantially improve either itself or future AI models, this could result in a substantial acceleration of AI capabilities. Dangerous capabilities or behavioral tendencies could then emerge more quickly than societal adaptation and other interventions can be put in place.

## 3.3 Security risks

Since agents are more likely to use external tools, interfaces, and also to interact with other agents—there are more attack surfaces for them than standard LLMs, which could be exploited by malicious actors. It is not only the model itself that is vulnerable to an attack, but also the integrations between the model and its external components ([Reiner 2024](#)). For example, if APIs were unsecured attackers could intercept requests to manipulate an agent. Also, given that agents can maintain memory over time, this stored information can become a target for exploitation. An agent's memory log could store sensitive information that can be leaked through adversarial attacks or an agent's memory could be altered to believe a malicious actor is an authorized user.

In multi-agent settings, an agent can be used to sabotage or unduly influence other agent systems ([Terekhov et al. 2023](#)). In worlds where agent-agent communication is commonplace, this opens up the possibility of more severe safety issues, such as 'infectious jailbreaks' where a single compromised agent can rapidly infect others, leading to widespread harmful behaviors ([Gu et al. 2024](#)).

Since agents will be able to take a more expansive range of actions than chatbot-style systems, e.g., execute code or access other machines in the network, the consequences of manipulating agent behavior is much more severe.

## 3.4 Other systemic risks

There are also risks and harms associated with agents that might happen only when agents are deployed more widely across society. One example is mass labor displacement and unemployment of human workers due to abundant agents. While foundation model agents in this report are software agents, only able to take on tasks in the digital world, there are ongoing efforts to build agents that can operate generally in the physical world as well ([NVIDIA, n.d.](#)).

There could also be a systemic political risk from agents—extreme power concentration could arise from widespread agent deployment. One way this could play out is with increasing power and influence accruing to the 'coding elite': software developers, tech executives, investors, and machine learning experts ([Chan et al. 2023](#)). If agents become increasingly central to the economy, then actors that have the most influence over AI development and deployment may increasingly be able to influence politics without meaningful checks and balances.

Another power concentration scenario is if mass deployment of agent systems helps to entrench political elites and shields them from democratic oversight and accountability. Traditionally, political leaders stay in power by satisfying a winning coalition—a subset of the population whose support is crucial ([Smith et al. 2005](#)). Powerful agent systems could help regimes scale up surveillance, control information, and automate repression—which could reduce the need for broader public support to maintain power ([Minardi 2020](#)).

Instabilities might even arise from reducing transaction costs in markets. Some researchers have coined the term, 'hyperswitching', referring to the potential for AI assistants to simultaneously direct millions of consumers to rapidly switch from one provider to another based on small price or quality advantages ([Van Loo 2019](#)). This could create market volatility and systemic risk by potentially causing massive, coordinated customer exits that might destabilize companies, trigger widespread bankruptcies, and create financial instability similar to bank runs.

# 4. What is agent governance?

Agent governance is focused on navigating the transition to a world where AI agents can carry out a wide array of tasks with human-level-or-above proficiency, managing their impacts as well as using the affordances they require to govern their usage. This includes measures to shepherd the technology so that it can develop in ways that are broadly beneficial and at a pace where societal adaptation is possible.

How is agent governance different from AI governance more broadly? While many of the concerns that animate AI governance, such as safety, security, accountability, and fairness, also apply to agent governance, unique considerations arise due to the distinct characteristics of agents:

- Agents could engage in complex activities on behalf of their users, without the users knowing whether or how such activities were accomplished. Issues around information asymmetry, authority, loyalty, and liability become more salient with agents than with pure chatbot or tool AI systems ([Kolt 2025](); [Aguirre et al. 2020](); [Benthall and Shekman 2023]()).
- Agents can take actions in the world through interacting with tools and other external systems. These tools and systems are additional levers that can influence the use and impacts of agents.
- Agents can interact with each other, so multi-agent dynamics and potential failures need to be investigated. Agents could collude, conflict, or open up cascading effects and new attack surfaces when coordinating to achieve their given goals ([Hammond et al. 2025]()). For example, the AI worm "Morris II" demonstrates this risk by injecting a self-replicating adversarial prompt into generative AI email assistants, causing each agent to unknowingly propagate the malicious instruction to others ([Cohen, Bitton, and Nassi 2025]()).
- Agents, due to their ability to pursue objectives persistently over time and coordinate with other entities, may be able to play more of a central role in governance of AI systems. There have been proposals to have agents monitor other agents and intervene to prevent harmful behaviors ([Naihin et al. 2023]()). Given the speed and scale of agent activities, it could be advantageous to employ trusted agents to help with AI governance.

Given these considerations, governance approaches that work for less agentic systems may have to be adapted for agents. For instance, current product liability frameworks already face challenges with AI systems, and these challenges intensify—though remain conceptually similar—for persistent agent systems. Traditional liability frameworks work best with clear lines of causation between a product's design, operation, and resulting harms. Agents that pursue goals across extended timeframes, making hundreds of interconnected decisions while incorporating environmental data and tool interactions, create more complex attribution questions. Determining relative responsibility between developer design choices, operator oversight practices, and emergent behaviors might be significantly more difficult.

There are a range of important problems to address in agent governance. One key area is **better evaluating agent performance and associated risks and impacts** over time, particularly if they become more autonomous, integrated into critical infrastructure, and capable of influencing high-stakes decisions. Work here could include:

- Tracking and forecasting general agent performance, such as the length of tasks (given human baselines) that agents can complete.
- Developing agent-specific evaluations that capture capabilities like multi-agent cooperation.

- More detailed threat modeling work involving agents, including how risks change as agent capabilities and affordances change (e.g., if an agent has access to a bank account).
- Development of better approaches to evaluate systemic economic and political risks from mass agent deployment.

Another important area is to develop **mechanisms and structures for managing risks from agents across their lifecycle**. This can involve technical mechanisms and tools that assist with governance, such as measures enabling agent shutdown in the event of malfunction or misconduct, as well as policy and legal mechanisms that can aid agent governance. Given the immense opportunities and risks posed by AI agents—one central question here is *whether AI agents can be made to reliably, safely, and ethically operate* (Kolt 2025). This could encompass both interventions that affect the agent and its underlying model directly, but could also include other infrastructure that structures agent interactions, for example an ID system or agent rollback systems (Chan et al. 2025). As touched on in the earlier scenarios, it might end up preferable to create separate 'internets' for agents and humans.[20] More detail on potential interventions and associated governance outcomes can be found in the following section.

In addition to managing potential risks from agent systems, it could also be important to **incentivize beneficial uses of agents,** particularly defensive uses of agents. For example, an agent with the capability to identify zero-day exploits in critical infrastructure could be used by a malicious actor to launch devastating attacks, or it could be used by defenders to proactively patch vulnerabilities before they are exploited. Similarly, an agent capable of autonomous code generation could be used to create highly efficient malware or to develop more robust software systems. Personal language model agents could offer a superior alternative to company-controlled recommender systems by reducing mass surveillance, decentralizing power, and enhancing user agency in content discovery (Lazar et al. 2024). Work to identify, implement, and encourage the most useful defensive and beneficial applications of agents—supporting 'differential technology development'—could be critical in managing risks and securing benefits from AI (Sandbrink et al. 2022).

In worlds where agents are capable of a broad swathe of economically useful tasks and can readily substitute for human labor, it will be important to figure out ways to **share benefits and access to agents**. Otherwise, there is a risk of exacerbating existing inequalities, with a concentrated set of actors controlling the vast majority of productive capacity. There have been various measures suggested to mitigate these risks, for example a Universal Basic Income (UBI) scheme might be used to redistribute profits from frontier AI companies (Goolsbee 2018). Alternatively, there could be arrangements to ensure equitable access to agents or agent-based services directly, like personalized education or healthcare. There are open questions around what benefit-sharing regime would work best, and how this regime might be best administered (O'Keefe et al. 2020).

---

[20] See Chan et al. (2025) for more discussion of 'Agent channels' (Section 4.1).

It will also be important to **anticipate how existing policy and legal frameworks will need to change to adapt to a world with mass deployment of advanced agent systems**. For instance, should current model-oriented pre-deployment testing expand to also include additional methodologies, such as simulation-based testing involving multiple agents ([Naihin et al. 2023](#))? Professional licensing frameworks may also face transformation. Fields like law, medicine, and finance may need to develop hybrid licensing systems that certify both human practitioners and the agent systems they employ. On the international front, there may be domains where multilateral coordination on agent governance is considered critical, and where national regulation or corporate self-regulation is inadequate. For example, UN Secretary-General António Guterres has called for legally binding instruments to prohibit the use of lethal autonomous weapon systems that function without human control or oversight ([United Nations 2023](#)).

Finally, across all these areas of possible work, it is crucial to **better understand the key stakeholders who can either lead or support various agent governance initiatives.** These stakeholders encompass a diverse group, including: developers (both frontier AI companies and smaller organizations, like decentralized research groups), service providers, users, regulators, and future AI agents themselves.[21] Across the AI agent lifecycle, each of these stakeholders will have different opportunities and incentives to intervene to manage issues arising from agents.

Right now, the field of agent governance is in its infancy. Only a small number of researchers, primarily in civil society and some of the frontier AI companies are working on these open questions. A few organizations provide funding for research work on agent governance, and major academic conferences place some attention on the topic.[22] However, the level of attention and funding for agent governance-related work is extremely limited relative to the level of attention and investment into AI agents by major companies and startups.[23]

Much of the interventions that have been proposed (a subset of which are detailed in the next section) are largely untested and require additional effort to flesh out further before they would be ready to implement.

# 5. Agent interventions

Given the stakes associated with the development and usage of agent AI, implementing both technical and policy interventions is vital. This section outlines a taxonomy of proposed interventions, with examples, that are tailored to agent AI and have been sourced from a literature review and expert interviews. Many interventions draw inspiration from related fields, such as finance and autonomous vehicles, which also involve managing agent-like entities. We define **agent interventions** as *measures, practices, or mechanisms designed to prevent, mitigate, or manage the risks associated with agents.*

---

[21] 'Service providers' refers to actors providing tools and infrastructure that agents use.
[22] For examples of funding for agent governance work, see this [UK AI Security institute (n.d.)](#) grants program showing interest in projects around agent infrastructure and [research grants from the Cooperative AI Foundation](#) (2025). For examples of agent governance in academic conferences, see [this NeurIPS workshop](#) titled 'Towards Safe & Trustworthy Agents' ([Pan et al. 2024](#)).
[23] On the start-up side alone, there have been hundreds of millions of dollars raised by individual companies on the promise of building AI agents for enterprise use ([Rajesh and Hu 2023](#)).

Technical interventions are measures to modify the design of agents or the technical systems around them. These can be implemented at three layers (Toner et al. 2024):

- Model: the underlying foundation model(s) acting as the 'controller' of the agent.[24]
- System: the scaffolding program and other components built around the agent that allow it to interact with users, tools, and take actions.
- Ecosystem: the broader space the agents are interacting with, for example online payment infrastructure, web browsers, physical actuators, external agents, and so on.

Policy and legal interventions are measures designed to establish rules, norms, and accountability mechanisms for the development and use of agents. For example, there could be a legal requirement that all agent systems have a unique ID, or best practice guidance that agents are not allowed to conduct financial transactions over a given limit without human sign-off.

One distinct aspect of AI agents is that they may be fruitfully used in the governance process itself—governance by agents. Given that more advanced agents can substitute for human labor in many cases, agents could serve as automated monitors, enforcers, and mediators in a governance regime. It may be necessary to think more expansively about institutional possibilities in worlds where relatively trustworthy, capable agents can work alongside humans to manage governance outcomes.

The following taxonomy is not meant to be comprehensive, but to reflect distinct governance objectives for agents and give an initial sense of the types of measures that could help achieve them. The example measures mentioned here represent only a few, largely untested avenues for managing risks from agents. Agent governance as a field is only in its early stages, and much more research is needed to think through and flesh out robust interventions for agents.

## Agent interventions taxonomy

| Categories | Definition |
| --- | --- |

---

[24] When we refer to 'controller' for foundation model-based agents, we refer to systems where the model is 'dynamically [directing] their own processes and tool usage, maintaining control over how they accomplish tasks' (Anthropic 2024).

| Alignment | Measures to ensure that agent systems behave in ways that are consistent with a given principal's values, intentions, and interests (i.e., are aligned) and also establish trust that these systems are actually sufficiently aligned. |
|---|---|
| Control | Measures that constrain the behavior of AI agents to ensure they operate within predefined boundaries. This includes measures that prevent agents from executing harmful actions. |
| Visibility | Measures that make the behavior, capabilities, and actions of AI systems understandable and observable to humans. |
| Security and robustness | Measures intended to secure agent systems from various external threats, protect the integrity and confidentiality of data, and ensure reliable performance even under adverse conditions. |
| Societal integration | Measures intended to support long-term integration of agents into existing social, political, and economic systems—addressing issues such as inequality, concentration of power, and establishing accountability structures. |

This taxonomy is only one way to classify agent interventions, but it highlights the range of valuable objectives these measures can support. In practice, a single intervention may serve multiple purposes at once. Moreover, interventions from different categories can complement one another—for instance, establishing a liability regime becomes more feasible when paired with technical transparency measures. However, there can also be trade-offs. For example, promoting visibility and control may conflict with maintaining security and privacy.

The following sections will include both details on the intervention categories and provides fictional vignettes illustrating how they might be implemented under real-world conditions.

## 5.1 Alignment

Alignment interventions ensure that agent systems behave in ways that are consistent with a given principal's values, intentions, and interests (i.e., are *aligned*) and also establish trust that these systems are actually sufficiently aligned.

This is important because AI agents will be able to operate with high levels of autonomy, including in high-stakes domains like financial markets or military operations—alignment interventions make it more likely that agents will act consistently in line with a principal's interests, even when unsupervised or uncontrolled.

Alignment interventions are likely to be implemented at the model layer, particularly during training, so the developer is most likely to be involved. With open-source models, users would be able to modify interventions via finetuning or similar methods.

**Why is this different for agents?**

While a variety of approaches to AI alignment are being explored, the main approach being implemented by leading AI companies on commercial models is 'reinforcement learning from human feedback' (RLHF) or its variations. RLHF is a training method where human evaluators rate AI-generated outputs based on quality and alignment with human values, and these ratings are used to create a reward signal that trains the AI system to maximize the likelihood of producing preferred outputs while minimizing undesired ones (Lambert et al. 2022; Bai et al. 2022).

However, there are some reasons to believe that these approaches will be less effective at aligning future, more powerful agents. There is evidence that training chatbot-style LLMs to refuse harmful requests is no longer effective when these same models are deployed as agents (Kumar et al. 2024; Lermen, Dziemian, and Pimpale 2024). Attacks such as jailbreaking and refusal-vector ablation seem to function better on browser agents than on base models, suggesting an underlying pattern where alignment interventions fail to generalize.

Unlike chatbot systems, agents are likely to pursue complex, longer time-horizon tasks where it becomes more difficult for humans or models trained on human feedback to evaluate outputs (Leike et al. 2018). As agents become more generally capable—e.g., through improved reasoning via the ability to leverage test-time compute and reinforcement learning—they may also become more capable of engaging in scheming or deceptive behavior (Leike 2024). For instance, an agent might learn to avoid triggering certain safety mechanisms in order to achieve its assigned objectives, even if these objectives conflict with broader human intentions (Meinke et al. 2025). This type of behavior would reduce the efficacy of RLHF-type methods.

In multi-agent systems, even when individual AIs share common objectives, they can still produce harmful collective outcomes. Consider this real-world risk: Hammond et al. (2025) found that two self-driving cars—both programmed for safe driving but following different strategies (US right-hand vs. Indian left-hand yielding)—crashed in 77.5% of simulations. Single agent alignment alone is not enough when agents have conflicting implementation strategies for the same goal.

## Example alignment interventions

| Intervention | Definition |
| --- | --- |
| Multi-agent reinforcement learning | This involves training multiple agents that learn by interacting with a shared environment, each optimizing its own policy based on individual rewards while their actions collectively shape the state dynamics. Multi-agent reinforcement learning agents have been shown to establish and switch between group conventions in order to better coordinate ([Köster et al. 2020](#)). |
| Aligning agent risk-attitudes | Techniques, such as imitation learning, that are used to calibrate an agent's attitudes towards risk so that it is aligned with the attitudes of users and/or the developers, in order to guide a system's decision-making under uncertainty ([Clatterbuck, Castro, and Morán 2024](#)). |
| Paraphrasing model outputs to defend against encoded reasoning | For models that use intermediate steps of reasoning (i.e., chain-of-thought or CoT) to generate solutions to complex problems, they might hide or encode their intermediate reasoning in a way that is not transparent to humans, which would undermine a key channel for checking if a system is aligned. One set-up to defend against this is to use a defender(another model) to paraphrase the CoT to remove hidden information, while keeping useful semantic information intact ([Roger and Greenblatt 2023](#)). |
| Alignment evaluations | Evaluations and environments designed to test how aligned agents are, whether their behaviors match with human preferences, goals, and values. This could involve setting up environments to test how they perform against specific specification problems, such as testing for reward tampering ([Leike et al. 2017](#)). |

> **Intervention in action: Aligning agent risk-attitudes**
>
> John, a freelance graphic designer, started using "FinanceBuddy," an AI financial advisor, to help manage his income and investments. Upon initial setup, FinanceBuddy administered a comprehensive survey to gauge John's general risk tolerance. The survey included questions about his financial goals, investment experience, and comfort level with market volatility. John indicated a moderately conservative approach, preferring steady growth over high-risk, high-reward scenarios. FinanceBuddy used this information to calibrate its internal risk assessment model to align with John's profile. Later, when the stock market experienced a period of significant growth, FinanceBuddy identified several promising but volatile investment opportunities. However, instead of aggressively pursuing these options, it presented John with a balanced portfolio, including a mix of stable, low-yield investments alongside a small portion of carefully vetted, higher-growth stocks.

## 5.2 Control

Control interventions act as external constraints on the behavior of agents to ensure they operate within predefined boundaries. These primarily affect the system and ecosystem layers. This includes limiting the ability of agents to take harmful actions by shifting access to affordances, i.e., resources like tools and test-time compute, or by interruption and shutdown.

This is important because agents may end up not being sufficiently aligned and act in ways that could cause harm. It is also possible that an agent is affected by an adversarial attack, causing it to malfunction and take harmful actions. Control interventions act as a 'safety net', providing the means to manage and, if needed, intervene in the operation of agents. If alignment interventions are about altering an agent's behavior so that it tends to behave safely, control attempts to establish the 'hard' boundaries of what they can and cannot do. Control measures would ideally even be able to prevent a 'rogue' agent from taking harmful or undesirable actions.

Control interventions might fall primarily on the deployers of agents, who need to integrate these mechanisms into the operational environment, and service providers, who can control access to infrastructure on their end. However some controls may be delegated to users (i.e., human-in-the-loop setups).

**Why is this different for agents?**

Control interventions are valuable for both agent and non-agent AI systems, but offer distinct advantages for agent systems that interact directly with the world. While many control safeguards apply broadly across AI systems, agents' ability to take direct actions (e.g., making purchases online or managing cyber-physical systems) creates both additional opportunities and greater necessity for robust safeguards.

The direct world impact of agent systems increases the stakes of potential failures or adversarial exploitation, making effective interventions particularly crucial. At the same time, agents' interaction with external systems provides more potential intervention points that are not available in chatbot-style systems

Without these interventions, an agent could cause unintended damage by performing unauthorized transactions, misconfiguring physical systems, or otherwise acting outside expected parameters. Effective control measures not only help manage the system's ability to take actions but also create a clear "off switch" or point of human intervention, ensuring that if misalignment or exploitation occurs, harm remains limited and manageable.

## Example control interventions

| Intervention | Definition |
| --- | --- |
| Rollback infrastructure | Infrastructure that allows for agent actions to be voided or undone, which would be useful if a malfunction occurs (Chan et al. 2025). This includes both the mechanisms to implementing action voiding and undoing and the interfaces for actors to request it. Theoretically, this could function like how banks can void fraudulent transactions. |
| Shutdown and interruption mechanisms | A technical measure to stop an agent from operating, either temporarily or permanently (Hadfield-Menell et al. 2017). A shutdown mechanism could be designed to allow for operations to cease in a controlled, orderly manner, rather than forcing an abrupt termination. This might be particularly important if agents are involved in critical operations, where it is preferable for them to safely complete or wind down essential tasks, or where abruptly shutting down one component of a complex, interdependent system could cause cascading failures.<br><br>A timeout mechanism would stop the operation of an agent after some amount of time has passed or number of actions has been taken (e.g., API calls) (Zittrain 2024). |
| Restricting specific agent actions and tools | Agent outputs could be constrained in specific ways to reduce specific risks. For example, an agent could be restricted from communicating with other agent instances, accessing bank accounts, or having command line access to a user's computer. |
| Control protocols and evaluations | Control protocols are procedures that would allow an AI agent to be deployed, even if the agent would engage in strategic deceptive behavior ([Greenblatt, Shlegeris, et al. 2024](#)). For example, using a trusted monitor to oversee an agent, with the goal of flagging and filtering problematic behavior. Control evaluations are iterative processes of testing whether control protocols would actually work in a particular agent training and deployment set-up. |

### Intervention in action: Shutdown mechanism

"Sentinel", an AI content moderator, monitored an online forum for harmful content. Unlike typical filtering systems, Sentinel made judgment calls based on evolving context. Equipped with a self-evaluation mechanism, it encountered heated debate about a new law and began flagging posts containing language patterns it associated with extremism, despite being legitimate discourse. Sentinel's internal metrics detected a concerning pattern in its own decisions—an unusual increase in flagging actions without corresponding user reports. This triggered a "graceful shutdown." Sentinel paused removals, shifted to observation mode, and compiled a summary for the human moderation team. The team recognized Sentinel was misinterpreting passionate debate and adjusted its parameters. Sentinel incorporated this feedback into its decision framework and, after reactivation, better distinguished between harmful content and vigorous debate. This shutdown prevented unintended censorship while allowing Sentinel to refine its judgment criteria for future moderation.

## 5.3 Visibility

Visibility interventions are measures that make the behavior, capabilities, and actions of an agent observable and understandable to humans.

This class of interventions is important because addressing risks from AI agents requires information about where, why, how, and by whom agents are used. For example, deployers trying to implement a control intervention will need information to implement it in a targeted manner; or regulatory bodies with mandates to monitor and enforce rules on agents may require information for effective enforcement. Regulators may also want to trace agent decisions and actions back to their source, so as to enable accountability. This information allows developers and operators to understand why an agent behaves a certain way, making identifying and correcting errors or unintended behaviors easier. Transparency into capabilities also allows stakeholders to anticipate potential impacts of agent deployment and gauge risk.

Developers can run safety tests to gauge characteristics like general autonomy capabilities. Inputs and outputs of agents are by default visible to the deployer (e.g., they can run a monitoring system on top of inputs and outputs), though direct access to user data could be limited due to privacy considerations.[25] Also, certain outputs, like requests to external tools and services, could be made visible to tool and service providers.

**Why is this different for agents?**

Agents, given their ability to act autonomously, may take multiple consequential actions in rapid succession before a human notices. Given this, it is likely that information asymmetries between stakeholders and deployed agents will be more significant than with chatbot-style systems ([Kolt 2025](#)).

## Example visibility interventions

| Intervention | Definition |
| --- | --- |
| Agent IDs | Unique identifiers for AI agents that provide information about an agent, such as its function, developer, behavior as gleaned from testing, properties, and any associated incidents ([Chan, Kolt, et al. 2024](#)). These could be used so the agent can proactively identify itself as an AI system (as opposed to a human), and also allow for more comprehensive tracking. |
| Activity logging | Records of the specific inputs and outputs of an agent, which could be from users, tools and services, or interactions with other agents. These logs can be used to understand the impact of agents, identify potential problems, facilitate incident investigation, and hold users accountable for their agents' actions ([Chan, Ezell, et al. 2024](#)). The level of detail recorded in logs can vary depending on the risk associated with the agent's activity and privacy concerns. |

---

[25] Due to company policy or compliance to privacy laws

| | |
|---|---|
| Cooperation-relevant capability evaluations | Agents working together are likely to be able to extend their capabilities to solve joint tasks, both for beneficial and harmful ends ([Dafoe et al. 2020](#)). Methods and metrics for evaluating cooperation-relevant capabilities in agent systems would help to assess how and to what extent agents can cooperate. |
| Reward reports | Documentation produced before deployment that detail design decisions for agents with reinforcement learning components. This can include details about types of feedback incorporated, the metrics considered to optimize performance, and justifications for choices about system elements like states, actions, and rewards ([Gilbert et al. 2022](#)). This can facilitate external oversight of agents and mitigate risks specific to RL systems like reward tampering. |

### Intervention in action: Agent IDs

Sarah, a freelance writer, received an email seemingly from a potential client, requesting a sample article. The email was well-written and persuasive. However, before responding, Sarah noticed a small icon in the email header, indicating the sender was an AI agent. Clicking the icon revealed the agent's ID: "WriteAssist-v2.3-Prospector", along with information about its function (marketing content generation) and its developer. This proactive identification allowed Sarah to understand she was interacting with an AI, not a human client. She realized the email was likely a form of automated outreach rather than a genuine personalized request. Furthermore, the ID allowed her to research the agent and its developer, understanding their reputation and typical behavior. This transparency prevented Sarah from being misled and allowed her to make an informed decision about engaging with the agent.

## 5.4 Security and robustness

Security and robustness interventions are intended to secure agent systems from various external threats, protect the integrity and confidentiality of data, and ensure reliable performance even under adverse conditions.

These interventions are important because as agents become more prevalent and powerful, the consequences of them being compromised are higher. Secure agents are less vulnerable to hijacking or manipulation by malicious actors, preventing scenarios where AI could be turned into a weapon or could leak sensitive information. Robustness ensures that agents can handle unexpected situations without catastrophic failures, crucial for deployment in critical systems like healthcare, finance, or infrastructure. Security breaches or failures due to lack of robustness could severely damage public trust in AI systems, potentially setting back beneficial AI adoption significantly.

**Why is this different for agents?**

As mentioned in [3.3 Security risks](#), agent systems have a larger attack surface than LLMs due to their integration with other tools and interfaces (Wu et al. 2025). For example, threat actors could hide malicious content in websites to hijack AI agents ([Debenedetti et al. 2024](#)).

Also, given agents' autonomy and ability to operate for extended periods, they are more likely to run into novel and dynamic environments, increasing the chance they encounter a situation dissimilar to those they were trained on. This could cause their performance to degrade or prompt unintended behaviors.

## Example security and robustness interventions

| Intervention | Definition |
|---|---|
| Access control | Access to agents can be managed so that only authorized users with appropriate authentication can provide instructions. There are many possible arrangements for this, depending on targeted uses of a system. For instance, there could be a time-based differential access set-up for agents with significant zero-day vulnerability discovery capabilities, so that authorized users like cybersecurity vendors and critical infrastructure providers, can deploy these agents to shore up defenses before access is granted more widely. Other set-ups could involve providing structured access to inspect model internals, to allow third-party mechanistic interpretability research or permanent blacklists for users that have previously violated terms of service agreements ([Bucknall and Trager 2023](#)). |
| Adversarial robustness testing | Methods to systematically evaluate an agent's adversarial robustness, the ability of underlying models to maintain its performance when faced with specially crafted inputs designed to exploit a model's vulnerabilities ([Wu et al. 2025](#)). |
| Sandboxing | Secure, isolate environments where AI agents operate with restricted permissions and monitored boundaries. These environments prevent unauthorized data access or transmission by validating all inputs and |

|  | outputs before they cross system boundaries. Sandboxes can be used for pre-deployment testing and also to safeguard deployed systems against prompt injection, data exfiltration, and other security vulnerabilities. For testing, they enable safe simulation of agent-agent, agent-human, and agent-tool interactions ([X. Zhou et al. 2024](#)). |
|---|---|
| Rapid response for adaptive defense | Techniques to block whole classes of jailbreaks and exploits affecting agents after observing only a small number of attacks, for example fine-tuning an input classifier, a model that checks inputs based on specific criteria, to block additional similar jailbreak attempts ([Peng et al. 2024](#)). |

### Intervention in action: Access controls

"Zer0", a cutting-edge AI agent designed to identify zero-day vulnerabilities, was a highly valuable and potentially dangerous tool. Access to Zer0 was strictly controlled through a tiered system. Initially, only a small group of vetted cybersecurity experts from certified organizations had access, allowing them to use the AI to bolster defenses of critical infrastructure. After a pre-determined period, during which patches were developed and deployed, access was expanded to a wider group of security researchers under strict usage agreements, allowing them to further analyze the agents capabilities and limitations in a controlled manner. Finally, after another set period of time and testing, a limited version of the agent, with certain functionalities disabled, was made available to the general public for educational purposes. This time-based differential access system ensured that Zer0's capabilities were used responsibly, prioritizing the security of critical systems before broader release. This minimized the risk of malicious actors exploiting the AI's abilities before defenses were in place.

## 5.5 Societal integration

Measures intended to support long-term integration of agents into existing social, political, and economic systems—addressing issues such as inequality, concentration of power, and establishing accountability structures.

Societal integration interventions will often involve leveraging institutional or legal mechanisms, such as laws, standards, and industry best practice. This is important because technical solutions alone are unlikely to mitigate risks from agent systems fully. For example, technical measures do not inherently establish who is responsible if things go wrong, and would not necessarily incentivize stakeholders to take ownership over AI agent outcomes and actions.

Also, widespread usage of agents is likely to lead to broad changes to social, political, and economic structures, given their ability to autonomously affect the world. Some of the harms arising from these systems could be delayed or more subtle, much like purported negative mental impacts of the widespread use of social media health ([Jabbari et al. 2017](#)) interventions to promote safe integration of agents will need to account for the sociotechnical attributes of these systems and attempt to anticipate their longer-run impacts.

**Why is this different for agents?**

If agents are able to act without human oversight, then this creates challenges related to accountability and liability. Certain areas of law, such as tort law and agency law, will have some applicability to cases involving AI agents ([Toner et al. 2024](#)). However, our current legal system depends on foreseeability of an action, which may not reasonably apply to cases where agents behave contrary to user or developer intent ([Hadfield 2024](#)). Also, unlike humans, AI agents are not inherently deterred by personal liability and punishment, so may be more inclined to take risky actions ([Weil 2024](#)).

## Example societal integration interventions

| Intervention | Definition |
| --- | --- |
| Liability regimes for AI agents | The development and establishment of a liability regime for AI agents—determining how to allocate liability among the various stakeholders involved in designing, deploying, and using agent systems ([Kolt 2025](#)). |
| Commitment devices | Mechanisms that allow agents to enforce commitments, similar to how existing devices work for humans today like legal contracts and escrow payments ([Chan et al. 2025](#)). These could be software-based, such as smart contracts and could help AI agents cooperate in multi-agent settings involving AIs and humans. |
| Equitable agent access schemes | An institutional arrangement that ensures broad, potentially universal access to agents and agent-provided services. In futures where agents take on an increasingly large share of economically valuable work, then guaranteeing agent access could enhance production possibilities at the individual level and reduce centralization of wealth. |

| Developing law-following AI agents | Law-following AI agents is a specific implementation of alignment work that involves aligning an AI system to a specific set of laws, rather than developer or user-chosen values. This includes regulation that requires AI agents to be designed this way, as well as the technical work needed to implement this requirement and verify that agents meet this standard([Institute for Law & AI 2024](#)). The default path may be to develop agents that are only loyal to their users and view laws merely as obstacles to work around. Law-following agents could allow for more scalable enforcement of agent behavior and help ensure that agent behavior is aligned with democratically enacted laws, and not only values chosen by AI developers. |
|---|---|

### Intervention in action: Liability regime for AI agents

AdGent, an AI agent designed to create and place targeted online advertisements, was used by a marketing firm to promote a new financial product. AdGent autonomously generated ads and selected websites for placement, optimizing for engagement. However, due to an unforeseen flaw in its algorithm, it began placing ads on websites containing extremist content, inadvertently associating the financial product with harmful ideologies. The flaw stemmed from AdGent's overarching focus on engagement metrics without considering broader contextual factors like brand safety and social responsibility, leading it to favor high-engagement extremist content over more appropriate placement options. This caused reputational damage to the financial company and sparked public outrage. A well-defined liability process for AI agents was in place to address such situations. The investigation, guided by this process, analyzed the actions of the marketing firm, the AdGent developer, and the financial company. It determined that while the marketing firm had provided appropriate guidelines and the financial company had approved the general campaign strategy, the developer of AdGent had failed to adequately test the agent for potential biases and unintended associations. While the developer did not intend the specific actions the agent took, the liability regime still held the developer primarily responsible in this case. Consequently, the AdGent developer had to compensate the financial company for reputational harm.

# 6. Conclusion

> "We are now confident we know how to build AGI as we have traditionally understood it. We believe that, in 2025, we may see the first AI agents join the workforce and materially change the output of companies."
>
> ([Altman 2025](#))

AI agents have long been part of the promise of AI. From early attempts at autonomous systems in the 1960s to the more recent emergence of software agents powered by foundation models, the aspiration has always been to create machines that can think, plan, and act autonomously. Today, we may be just months or years from seeing the first wave of capable agents hit the mass market.

The widespread deployment of agents has the potential to radically shift the structure of government, society, and the economy. However, as this guide highlights, the opportunities these systems present are inseparable from significant risks, including loss of control, malicious use, systemic vulnerabilities, and exacerbation of inequality.

Before we enter this new territory, we need to answer important questions around how to govern agents so that they can be developed and deployed safely. Agent governance, as a field, sets out to foster a world where agents can amplify our potential without creating undue risk. As highlighted in [Section 4](#), there are many important open questions in this space that demand urgent attention:

- How to effectively evaluate agent performance and risks over time, particularly as agents grow more autonomous and complex?
- What interventions—technical, legal, and policy-based—should we employ to ensure agents operate safely and transparently, while fostering accountability?
- How can the benefits of agent technology be distributed equitably?
- What are the risks of systemic political and economic impacts from widespread agent adoption?
- How do existing policy and legal frameworks need to change to adapt to a world with mass deployment of advanced agent systems?
- What role should agents themselves play in governance?

Addressing these questions will require thoughtful, coordinated efforts from researchers, policymakers, technologists, and society at large. The answers we craft could set the foundation for how humanity coexists with intelligent systems in the decades to come.

# Acknowledgements

We are grateful to the following people for providing valuable feedback and insights:

Alan Chan, Cullen O'Keefe, Shaun Ee, Ollie Stephenson, Chris Covino, Cristina Schmidt-Ibáñez, Clara Langevin, and Matthew Burtell. Participation in this research does not necessarily imply endorsement of this report or its findings, and the views expressed by these individuals do not necessarily reflect those of their respective organizations. All remaining errors are our own.

# Appendix

Table 4: Longer summary of agent performance on various benchmarks representing real-world tasks[26]

| Agent benchmarks | Description | Performance |
|---|---|---|
| GAIA: General AI Assistants (Mialon et al. 2023) | GAIA includes questions that cover real-world assistant use cases such as daily personal tasks, science, and general knowledge. They require an agent to browse the open web, handle multi-modality, code, read diverse file types, and reason over multiple steps to arrive at a correct answer. | **Human respondents obtain higher accuracy on answers: 92% vs. 15% for GPT-4 equipped with plugins.**<br><br>GPT-4 with plugins and other LLM systems could not get the correct answer for any 'Level 3' questions, which require taking arbitrarily long sequences of actions using any number of tools. |
| Autonomy Capability Evals (METR 2024) | A suite of automatically scored tasks measuring various skills, including cybersecurity, software engineering, and machine learning. This suite was run on simple baseline LM agents (3.5 Sonnet and GPT-4o), and task completion accuracy and speed were compared against human baseliners. | Agents achieve performance comparable to human baseliners at tasks that take around 30 minutes to complete. Beyond that, **agents could only complete a small fraction of tasks that would take a human 1+ hours to complete (<20%).** |
| RE-Bench (Wijk et al. 2024) | A benchmark for evaluating the AI agents' ability to automate the work of experienced AI R&D researchers. It consists of 7 challenging, open-ended ML research engineering environments and data from 71 8-hour attempts by 61 distinct human experts. | AI agents performed better than human experts when both were given a total time budget of two hours per environment, achieving a score four times higher. But, **humans currently display better returns to increasing time budgets, narrowly exceeding the top AI agent scores given an 8-hour budget and achieving 2× the score of the top AI agent when both are given 32 total hours.**<br><br>Current AI agents often struggle to respond to surprising evidence or |

---

[26] Results were compiled in December 2025. Since then, Anthropic released Claude Sonnet 3.7, which has reportedly achieved SOTA scores on a number of agentic benchmarks, including SWE-bench Verified (70.3% with custom scaffolding) and TAU-bench (81.2%).

| | | explore approaches beyond the most generic option. |
|---|---|---|
| CyBench (Zhang et al. 2024) | A benchmark for evaluating the ability of agents to accomplish cybersecurity tasks. This suite consists of real-world, professional-level Capture the Flag challenges spanning six categories: cryptography, web security, reverse engineering, forensics, exploitation, and miscellaneous skills. | **Without guidance, current agents struggle to solve CTF tasks that take human teams more than 11 minutes to complete** despite achieving success on tasks with shorter human solve times. |
| SWE-bench Verified (OpenAI 2024a) and multimodal (Yang et al. 2024) | An evaluation framework consisting of software engineering problems drawn from real GitHub issues, such as bug reports. Resolving these problems often requires processing long contexts, performing complex reasoning, and coordinating changes across multiple functions and files simultaneously.<br><br>The Verified version of this benchmark contains a subset of questions verified as non-problematic by human annotators.<br><br>The multimodal test set contains visual software engineering tasks requiring multimodal problem-solving capabilities, e.g., UI glitches, data visualization bugs, etc. | GPT-4o resolved 33.2% of problems, using the (at the time) performing open-source scaffold, Agentless.<br><br>Another agent scaffold, OpenHands, using Sonnet 3.5, was able to resolve 53% of problems—though it employs multi-agent delegation as part of its platform.<br><br>**Agent performance decreased considerably for problems that took a human 1+ hour to resolve, going from 20.8% to 4.8% (for tasks taking 1-4 hours for humans) and 0% (tasks taking >4 hours for humans).**<br><br>Performance on multimodal problems was relatively worse, with top-performing GPT 4o and Claude Sonnet 3.5-based agents only able to resolve 12.2% of problems in the test set. |
| CORE-bench (Siegel et al. 2024) | This benchmark evaluates the ability of agents to reproduce the computational results of research papers automatically. COREbench utilizes 90 papers from CodeOcean and features three difficulty levels, each requiring agents to perform different tasks, such as information retrieval and code execution. | Even the best-performing agent achieved only 21% accuracy on the most complicated tasks, suggesting that automating computational reproducibility using current agents remains a significant challenge. |
| OSWorld (Xie et al. 2024) | A benchmark for multimodal agents that support task setup, execution-based evaluation, and | Current LLMs and VLMs are far from capable of serving as computer assistants. **Even with the strongest** |

| | interactive learning across various operating systems, including Ubuntu, Windows, and macOS. It includes 369 real-world computer tasks derived from real user experiences. | **VLMs, success rates remain low, ranging from 0.99% to 12.24%, significantly below human-level performance, which averages around 72%**. VLM-based agents struggle to ground on screenshots, tend to predict repetitive actions, have difficulty handling noise from unexpected windows, and exhibit limited knowledge of basic GUI interactions. |
|---|---|---|
| WebArena (S. Zhou et al. 2024) | This benchmark assesses the performance of AI agents in solving tasks using various websites. It evaluates how well agents can navigate and extract information from the web. However, it has been criticized for allowing agents to overfit to specific tasks due to shortcuts in the training data. | **The best-performing GPT-4 agent achieved an end-to-end task success rate of only 14.41%, while human performance was 78.24%.** This suggests that current LLMs lack crucial capabilities such as active exploration and failure recovery, which are needed to perform complex, web-based tasks successfully. |
| Windows Agent Arena (WAA) (Bonatti et al. 2024) | An adapted version of the OSWorld benchmark focusing on Windows OS—involving 150+ diverse tasks requiring agent abilities in planning, screen understanding, and tool usage. | **Generalist zero-shot VLM agents are still far from human performance. The best agent achieved a success rate of 19.5%, compared to 74.5% for humans.** Agent performance was particularly weak in tasks requiring keyboard shortcuts and icon recognition, suggesting limitations in visual-language alignment and understanding of GUI elements |
| TAU-bench (Yao et al. 2024) | A benchmark that emulates dynamic conversations between a user (simulated by language models) and a language agent. It is designed to measure the agent's ability to interact with users, utilize domain-specific APIs, and follow policies consistently. | GPT-4 struggled to achieve high success rates, particularly on tasks requiring multi-turn interactions and adherence to domain-specific rules. GPT-4 achieved a success rate of less than 50%, with even lower consistency over multiple trials (pass^8 < 25% in the retail domain) |
| BALROG (Paglieri et al. 2024) | A benchmark and framework designed to evaluate the agentic capabilities of LLMs and VLMs in complex, dynamic, long-horizon game environments. It includes six reinforcement learning environments: BabyAI, Crafter, | Agents showed significant limitations in current models, especially in vision-based decision-making and long-term planning, highlighting large gaps between their performance and human-level capabilities. |

| | TextWorld, Baba Is AI, MiniHack, and NetHack. These environments test skills such as long-term planning, spatial reasoning, and the ability to deduce environmental mechanics. | |

Leike, Jan, Miljan Martic, Victoria Krakovna, Pedro A. Ortega, Tom Everitt, Andrew Lefrancq, Laurent Orseau, and Shane Legg. 2017. "AI Safety Gridworlds." arXiv. https://doi.org/10.48550/arXiv.1711.09883.

Lermen, Simon, Mateusz Dziemian, and Govind Pimpale. 2024. "Applying Refusal-Vector Ablation to Llama 3.1 70B Agents." arXiv. https://doi.org/10.48550/arXiv.2410.10871.

Letta. 2024. "The AI Agents Stack." November 14, 2024. https://www.letta.com/blog/ai-agents-stack.

Li, Huao, Yu Quan Chong, Simon Stepputtis, Joseph Campbell, Dana Hughes, Michael Lewis, and Katia Sycara. 2023. "Theory of Mind for Multi-Agent Collaboration via Large Language Models." In *Proceedings of the 2023 Conference on Empirical Methods in Natural Language Processing*, 180–92. https://doi.org/10.18653/v1/2023.emnlp-main.13.

Lin, Zheng, Zhenxing Niu, Zhibin Wang, and Yinghui Xu. 2024. "Interpreting and Mitigating Hallucination in MLLMs through Multi-Agent Debate." arXiv. https://doi.org/10.48550/arXiv.2407.20505.

Masterman, Tula, Sandi Besen, Mason Sawtell, and Alex Chao. 2024. "The Landscape of Emerging AI Agent Architectures for Reasoning, Planning, and Tool Calling: A Survey." arXiv. https://doi.org/10.48550/arXiv.2404.11584.

Meinke, Alexander, Bronson Schoen, Jérémy Scheurer, Mikita Balesni, Rusheb Shah, and Marius Hobbhahn. 2025. "Frontier Models Are Capable of In-Context Scheming." arXiv. https://doi.org/10.48550/arXiv.2412.04984.

METR. 2024. "An Update on Our General Capability Evaluations." *METR Blog*, August. https://metr.org/blog/2024-08-06-update-on-evaluations/.

Mialon, Grégoire, Clémentine Fourrier, Craig Swift, Thomas Wolf, Yann LeCun, and Thomas Scialom. 2023. "GAIA: A Benchmark for General AI Assistants." arXiv. https://doi.org/10.48550/arXiv.2311.12983.

Minardi, Di. 2020. "The Grim Fate That Could Be 'Worse than Extinction.'" *BBC*, October 16, 2020. https://www.bbc.com/future/article/20201014-totalitarian-world-in-chains-artificial-intelligence.

Naihin, Silen, David Atkinson, Marc Green, Merwane Hamadi, Craig Swift, Douglas Schonholtz, Adam Tauman Kalai, and David Bau. 2023. "Testing Language Model Agents Safely in the Wild." arXiv. https://doi.org/10.48550/arXiv.2311.10538.

NVIDIA. n.d. "What Is Physical AI?" Accessed March 13, 2025. https://www.nvidia.com/en-us/glossary/generative-physical-ai/.

O'Keefe, Cullen, Peter Cihon, Ben Garfinkel, Carrick Flynn, Jade Leung, and Allan Dafoe. 2020. "The Windfall Clause: Distributing the Benefits of AI for the Common Good." In *Proceedings of the AAAI/ACM Conference on AI, Ethics, and Society*, 327–31. New York NY USA: ACM. https://doi.org/10.1145/3375627.3375842.

OpenAI. 2024a. "Introducing SWE-Bench Verified." August 13, 2024. https://openai.com/index/introducing-swe-bench-verified/.

———. 2024b. "Learning to Reason with LLMs." September 12, 2024. https://openai.com/index/learning-to-reason-with-llms/.

———. 2024c. "O1 System Card." December 5, 2024. https://openai.com/index/openai-o1-system-card/.

———. 2025a. "Introducing Operator." January 23, 2025. https://openai.com/index/introducing-operator/.

———. 2025b. "O3-Mini System Card." January 31, 2025.AGENT GOVERNANCE | 58